\pdfoutput=1
\documentclass[appendixfloats,tighten,iop,numberedappendix]{emulateapj}
\usepackage{hyperref}

\begin{document}
\title{On the Performance of Quasar Reverberation Mapping in the Era of 
Time-Domain Photometric Surveys}
\author{Doron Chelouche\altaffilmark{1},  Ohad Shemmer\altaffilmark{2}, Gabriel I. Cotlier\altaffilmark{1}, Aaron J. Barth\altaffilmark{3}, and Stephen E. Rafter\altaffilmark{1,4}}
\altaffiltext{1} {Department of Physics, Faculty of Natural Sciences, University of Haifa, Haifa 31905, Israel; doron@sci.haifa.ac.il}
\altaffiltext{2}{Department of Physics, University of North Texas, Denton, TX 76203, USA; ohad@unt.edu}
\altaffiltext{3}{Department of Physics and Astronomy, 4129 Frederick Reines Hall, University of California, Irvine, CA 92697, USA; barth@uci.edu}
\altaffiltext{4}{Physics Department, Technion, Haifa 32000, Israel; e-mail: rafter@physics.technion.ac.il}
\shortauthors{Chelouche et al.}
\shorttitle{Reverberation Mapping of Quasars in the Era of Photometric Surveys}

\begin{abstract}

We quantitatively assess, by means of comprehensive numerical simulations, the ability of broad-band photometric surveys to recover the broad emission line region (BLR) size in quasars under various observing conditions and for a wide range of object properties. Focusing on the general characteristics of the Large Synoptic Survey Telescope (LSST), we find that the slope of the size-luminosity relation for the BLR in quasars can be determined with unprecedented accuracy, of order a few percent, over a broad luminosity range and out to $z\sim 3$. In particular, major emission lines for which the BLR size can be reliably measured with LSST include H$\alpha$, \ion{Mg}{2}\,$\lambda 2799$, \ion{C}{3}]\,$\lambda 1909$,  \ion{C}{4}\,$\lambda 1549$,  and Ly$\alpha$, amounting to a total of $\gtrsim 10^5$ time-delay measurements for all transitions. Combined with an estimate for the emission line velocity dispersion, upcoming photometric surveys will facilitate the estimation of black hole masses in AGN over a broad range of luminosities and redshifts, allow for refined calibrations of BLR size-luminosity-redshift relations in different transitions, as well as lead to more reliable cross-calibration with other black hole mass estimation techniques.

\end{abstract}


\section{Introduction}

Mass estimation of supermassive black holes (SMBHs) in active galactic nuclei (AGN) relies on locally-established relations between the SMBH mass and various source observables, such as the luminosity and the velocity dispersion of the broad emission line region, BLR \citep{kas00,pet04,vp06,ben09,den09}. While the adequacy of such relations has been demonstrated for local samples of objects where several independent means for estimating BH masses exist (e.g., via the reverberation mapping technique and the stellar velocity dispersion in the inner regions of the host; \citealt{on04}), it is not clear that this approach is warranted also at higher $z$, where one probes earlier cosmic times, and is sensitive only to the most luminous sources at those epochs \citep{net03}. Further, as quasar\footnote{The terms AGN and quasars are used here interchangeably.} observables in a given spectral band are redshift-dependent, it is not clear that inter-calibration of various phenomenological relations (e.g., between luminosity, the particular emission line probed, and SMBH mass), even if justified, is free of biases \citep{den09}. In this respect, the challenge in SMBH mass estimates at high-$z$ is akin to the cosmological distance scale problem, with more data and better control of systematics required to place them on firmer ground. 

With upcoming (photometric) surveys that will monitor a fair fraction of the sky to unprecedented depth and photometric accuracy, and with good cadence, the time-domain field of quasar study, and specifically reverberation mapping of the BLR, is expected to undergo a major revision \citep{cd11}. As has been demonstrated in several previous works \citep{ha11,cd11,cd12,c13,cz13,ed12,po12,po13,zu13}, while individual emission lines cannot be resolved using photometric means, their signal can be recovered at the light curve level \citep[see also][]{raf13}, and reverberation mapping is, in principle, possible. 

In this work we build upon the formalism of \citet{cz13}, and provide more realistic benchmarks for reverberation mapping using broadband photometric surveys, having in mind the characteristics of the {\it Large Synoptic Survey Telescope} (LSST)\footnote{\url{http://www.lsst.org/lsst/scibook}} experiment. Specifically, we are interested in the ability of photometric campaigns in determining BLR-associated time delays in individual sources as well as statistically, in sub-samples of sources. This paper is organized as follows: section 2 describes the model used here for constructing mock AGN light curves in different bands, as well as the analysis technique employed to recover the input line-to-continuum time-delays. Section 3 provides benchmarks for the measurement of the line-to-continuum time-delay under various assumptions concerning the sampling, redshift determination accuracy, filter choice, object properties, and the priors used in the analysis. The discussion, with particular emphasis on LSST-enabled science, follows in section 4, and the summary in section 5.

\section{Light Curve Simulations and their Analysis}

Below we specify our approach for simulating quasar light curves. For demonstrative purposes, we use the characteristics of the LSST and standard cosmology\footnote{$(h,\Omega_M,\Omega_\Lambda)=(0.7,0.3,0.7)$}. 

\begin{figure}
\epsscale{1.2}
\plotone{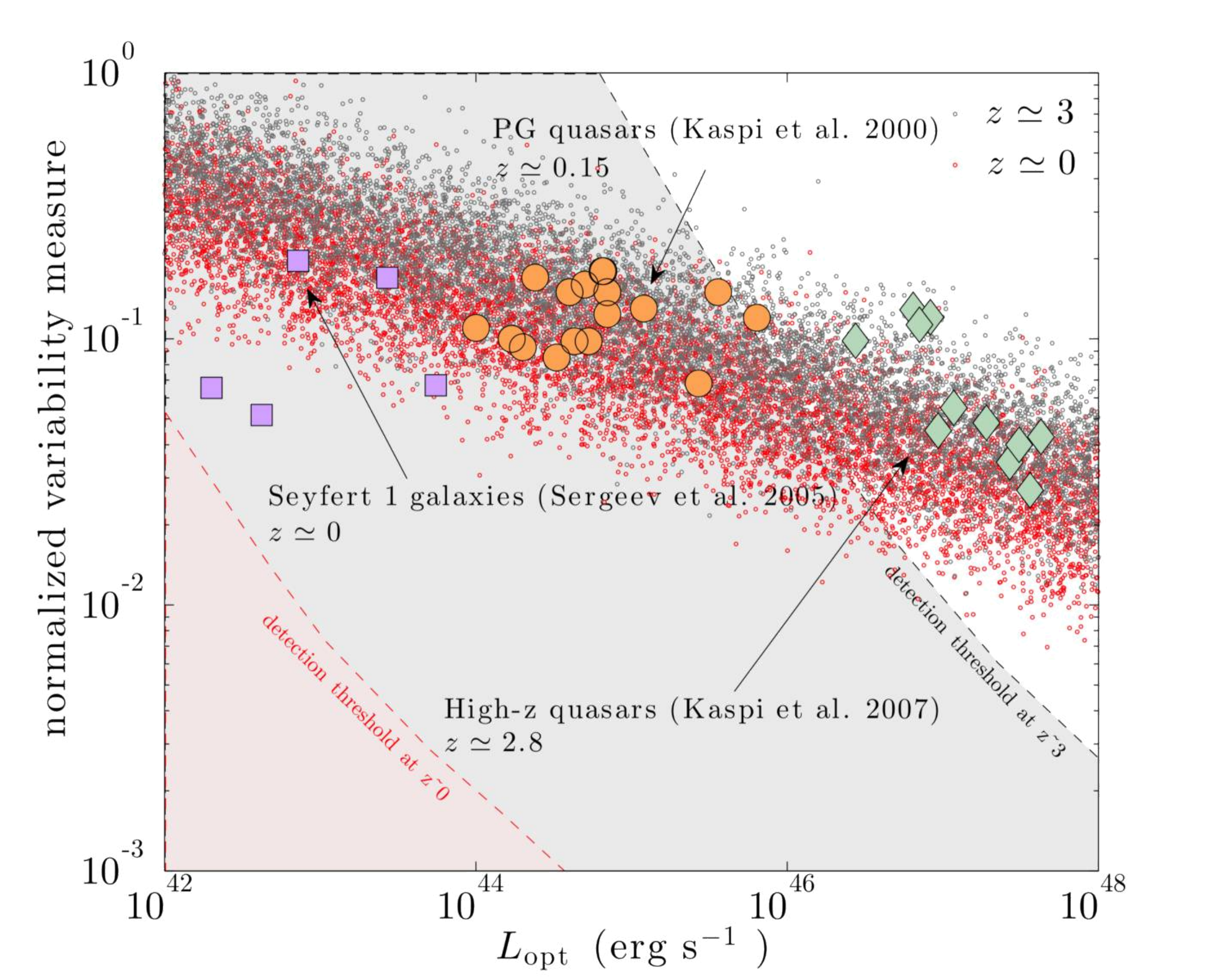}
\caption{Normalized variability measure for simulated quasars, $\sigma_f^r$, based on Eq. \ref{ep}, for two redshift bins (see legend), and as a function of object luminosity shows a trend of decreasing value with luminosity. Redshift effects are relatively minor (a $\gamma=-1$ was assumed here for demonstrative purposes). Also shown are data for low- and high-luminosity AGN, as well as detection thresholds based on single LSST exposures of the field. Low-luminosity AGN from the \citet{ser05} sample may have their normalized variabilities underestimated due to dilution of the AGN signal by the host galaxy emission. Also, extremely radio-loud objects (blazars) are overrepresented in the sample of \citet{kas07} with respect to the local quasar population.}
\label{var}
\end{figure}

\subsection{Continuum light curves}

We model the continuum light curve, $f_c$, using the method of \citet{tk95} so that its Fourier transform is of a powerlaw form, $\tilde{f}_c(\omega)\sim \omega^\gamma$, where we choose $\gamma=-1.3$ to be the frequency powerlaw slope \citep[note that our definition for $\gamma$ differs from that used in, e.g., ][ who report instead slopes for the power spectrum]{giv99}. Our choice of $\gamma$ results in a power spectrum that is somewhat redder than the one which characterizes continuous first-order auto-regressive [CAR(1)] processes, for which $\gamma=-1$ \citep{kb09}. This is in better agreement with the slopes found by high-cadence observations of AGN using {\it Kepler} data \citep{mu11}, and may be more characteristic of the variability of high-$z$ sources given the cadence and lifetime of the LSST experiment \citep{lsst}. We explore departures from our chosen value of $\gamma$ in section 3.

\subsubsection{The light curve variance}

The normalized variability measure (i.e., the standard deviation of the light curve normalized by the mean flux level, and ignoring measurement uncertainties) in some filter $j$, which is characterized by a central wavelength, $\lambda_j$, is $\sigma_f^j$, and depends, in principle, on the quasar luminosity, $L$, its redshift, $z$, and the wavelength range probed. Other quasar properties, such as radio loudness, black hole mass, and its Eddington ratio may also be important but are not taken into account in the present treatment of the problem. In particular, we do not consider blazars or sources where the jet contributes significantly to the variability. 

Our model for $\sigma_f^j$ is of the form 
\begin{equation}
\sigma_f^j(L,\lambda,z)=\sigma_f^r(z=0) L_{\rm opt,45}^\beta \left ( \frac{\lambda_j}{\lambda_r} \right )^\zeta (1+z)^\epsilon,
\label{ep}
\end{equation}
 where $\sigma_f^r(z=0)$ is the rest frame normalized variability, $L_{\rm opt,45}=L_{\rm opt}/10^{45}\,{\rm erg~s^{-1}}$ where $L_{\rm opt}$ is the rest-frame optical luminosity of the AGN, and $\beta,\zeta,\epsilon$ are powerlaw indices whose values are discussed below, and are assumed to be independent of redshift and AGN characteristics. 
 
Motivated by the fact that the variance of the light curve is $\int_{\omega_{\rm min}}^{\omega_{\rm max}} \vert \tilde{f}(\omega) \vert^2 d\omega\sim \omega_{\rm min}^{2\gamma+1}$ (here we assumed red-noise spectra and $\omega_{\rm max}\to\infty$), we note that, unless the physical bound on $\omega_{\rm min}$ is observable (e.g., when a spectral break lies in the frequency range probed by the time series), the reduced variability measure of the light curve depends on the duration of the experiment, $\Delta t$, so that $\sigma_f^j\sim \Delta t^{ -(\gamma+1/2)}$. For an experiment of a fixed duration over the entire sky (e.g., LSST) we get that $\sigma_f^j\sim (1+z)^{\gamma+1/2}$ because of this time dilation effect alone. Nevertheless, a further effect determines the final value of $\epsilon$: observationally, it is well established that longer (rest) wavelength data have smaller variations such that their reduced variability measure $\sim \lambda^{-0.8}$ hence $\zeta=-0.8$ \citep{ga99,giv99,m11}. Therefore, observing quasars over a broad range of redshifts, using a particular band, would lead to an independent redshift dependence of the form $\sigma_f^j\sim (1+z)^{0.8}$. Combining the two sources for redshift dependence, we obtain that, for $-1.3<\gamma <-1$, $0<\epsilon<0.3$. In this work we do not consider the possibility of quasar evolution, i.e., an intrinsic dependence of quasar variability on the cosmological epoch, an assumption which appears to be consistent with the data \citep{m11}. If present, this could imply enhanced (suppressed) high-$z$ variability of quasars than assumed here, thereby facilitating (restricting) reverberation mapping.

The luminosity dependence of the normalized variability measure is determined by searching for a reasonable agreement between the model predictions (equation \ref{ep}), and the combined variability dataset for Seyfert galaxies \citep{ser05} and Palomar-Green (PG) quasars \citep{giv99,kas00} at low $z$. The model is then scaled up to higher $z$ using equation \ref{ep}, and its predictions are checked against data for luminous high-$z$ objects \citep{kas07}, which cover the relevant timescales. The normalization constant, $\sigma_f^r(z=0)$ and $\beta$ are then simultaneously determined yet we note that the solution may not be unique, and we make no attempt to cover the plausible solution phase space in this work. We find that $\beta\sim -0.2$ and $\sigma_f^r(z=0)\sim 0.2$ provide an adequate description for the median reduced variability measure of all data sets. Nevertheless, we caution that (a) the normalized variability measure for some Seyfert galaxies is somewhat underestimated due to host contamination in the \citet{ser05} dataset \citep{cak07}, and that (b) some of the high-$z$ objects in the \citet{kas07} sample are extremely radio-loud objects, blazars, whose variability may not be characteristic of the bulk of the high-$z$ radio-quiet population. 

Lastly, individual objects show considerable scatter around the typical (median) normalized variability measure at a given luminosity range.  We find that a model where equation \ref{ep} is multiplied by a factor $e^\delta$, where $\delta$ is a Gaussian random variable with a standard deviation of $0.4$ and a zero mean, qualitatively agrees with the observations (see Fig. 1). Overall, our model provides a fair statistical description of quasar variability, which is in line with the qualitative statistical framework considered in this work. However, it is not designed to account for the full richness of quasar variability (e.g., with a possible dependence also on the BH mass, or the Eddington ratio), nor for the particular properties of extreme AGN types (e.g., the narrow line objects), and the possibility of truly non-stationary light curve behavior.

\subsubsection{Time delays}

It is known that the continua light curves of quasars in different bands are highly correlated \citep{ul97}. Nevertheless, there are a few examples of low-luminosity Seyfert 1 galaxies showing discernible lags between the continuum light curves in different bands \citep{col98,col01,cz13,c13}. These are consistent with a physical picture in which the inner accretion disk irradiates its outer parts resulting in longer wavelength reprocessed emission lagging the short wavelength one (see however \citealt{kor01} for a different interpretation of the data). We note, however, that such time delays have not been established for high-luminosity sources, nor in high-$z$ quasars.

In cases where continuum transfer effects across the accretion flow are relevant, the light curve in some band $k$, may be obtained from the light curve in band $j$ by means of a convolution with an appropriate transfer function, $\psi_c(t)$. For simplicity, we shall assume a rectangular function such that $\psi_c(t)=1/2\tau_{jk}$, and is defined in the range $t\in [0,2\tau_{jk}]$, where $\tau_{jk}$ is the time delay between the bands (for more theoretically motivated transfer functions see \citealt{cak07}, \citealt{c13}, and references therein). We use theoretically expected time delays from standard thin accretion-disk theory, which seem to be supported by the current data, such that
\begin{equation}
\tau_{jk}= 7L_{\rm opt,45}^{0.5}(1+z)^{-1/3} \left [ \frac{\lambda_k^{4/3}-\lambda_j^{4/3}}{5500{\rm \AA}^{4/3}} \right ]\,{\rm days}
\label{tjk}
\end{equation}
where we implicitly assumed that $\lambda_k>\lambda_j$, and normalized according to \citet{c13}. The redshift dependence  results from the combined effect of cosmological time dilation, and the smaller sizes of the accretion disk probed by shorter wavelength emission that enters the optical bands at higher $z$. The continuum light curve in band $k$ is then set once $f_c^j$ is determined by
\begin{equation}
f_c^k=f_c^j*\psi(\tau_{jk}),
\end{equation}
where '$*$' denotes convolution. If \citet{ss73} disks provide a fair description of the accretion physics in AGN, then this relation is expected to hold except for objects with the lowest/highest accretion rates. In those physical regimes, advection becomes relevant \citep{ab88,ny94}.

\subsection{Line and non-Ionizing Continuum Emission}

\begin{table}
{\footnotesize
\begin{center}
\caption{Emission Components}
\begin{tabular}{llll}
\tableline
 & $\lambda_i$ & $W_0$ & $\tau$  \\
ID  &  $[{\rm \AA} ]$ & $[{\rm \AA} ]$ & $[$days$]$ \\          
\tableline
(1) Ly$\epsilon$ & 941  & 4.7 & 42 \\
(2) \ion{C}{3} & 977 &  6.6 & 50 \\
(3) Ly$\beta$ & 1033  & 9.8 & 42 \\
(4) \ion{Fe}{3} & 1117  & 3.7 & 50 \\
(5) Ly $\alpha$ & 	1216  & 93 & 42 \\
(6) \ion{N}{5}&1240&1&18\\
(7) \ion{Si}{4}&	1397		&8	&50 \\
(8) \ion{C}{4}&	1549		&24	&40\\
(9) \ion{C}{3}$]$&	1909		&21	&120 \\
(10) \ion{Fe}{3}&	2077		&2.5	&50\\
(11) \ion{Fe}{2}&	2324		&3.6	&50 \\
(12) \ion{Mg}{2}&	2799		&32	&70 \\
(13) \ion{Fe}{2}&	2964		&5	&50 \\
(14) \ion{He}{1}&	3189		&1	&100 \\
(15) \ion{Fe}{2}&	3498		&1.4	&50 \\
(16) H$\delta$&	4103&		5&	36 \\
(17) \ion{Fe}{2}&	4160		&1	&50 \\
(18) H$\gamma$&	4342&		13	& 60 \\
(19) \ion{Fe}{2}&	4564		&20	&200 \\
(20) \ion{He}{2} & 4687 & 1 & 40 \\
(21) H$\beta$&	4863		&46	&100 \\
(22) \ion{Fe}{2}&	5305		&22	&200 \\
(23) \ion{He}{1}&	5877&		5&	100 \\
(24) H$\alpha$&	6565		&195	&130 \\
(25) \ion{He}{1}&	7067&		3&	100 \\
(26) \ion{O}{1}&	8457&		10&	200 \\
(27) \ion{Fe}{2}&	9202&		4&	200 \\
(28) Pa$\epsilon$&	9545	&	7&	150 \\
(29) Pa$\delta$&	10049		&21	&150 \\
(30) \ion{He}{1}&	10830		&36	&100 \\
(31) Pa$\gamma$&	10941		&7	&150 \\
\tableline
(32) Balmer cont. &	3400-4000$^\dagger$&	100&	30 \\
(33) Paschen cont.&	6500	-8300$^\dagger$&	300$^\ddag$&	150 \\
(34) Hot dust	&10000-20000$^\dagger$	&10000	&500 \\
\tableline
\end{tabular}
\end{center}
BLR emission features (i.e., lines, line blends, and continua) included in this work, with their parameters ($W_0$ and $\tau$) given in the restframe; see table 2 in \citet{dvb01} for wavelength definitions. Quoted lags correspond to an $L_{\rm opt,45}=1$ quasar (see text). \\
$^\dagger$ Approximate wavelength are used. For the recombination continua, allowance is made for the combined contribution of higher-order emission lines red-ward of the edge ionization energy. \\
$^\ddag$ The flux in this component is poorly constrained and $W_0$ is inspired by the theoretical calculations of \citet{kk81,kor01}.}
\end{table}

\begin{figure*}
\epsscale{1.22}
\plotone{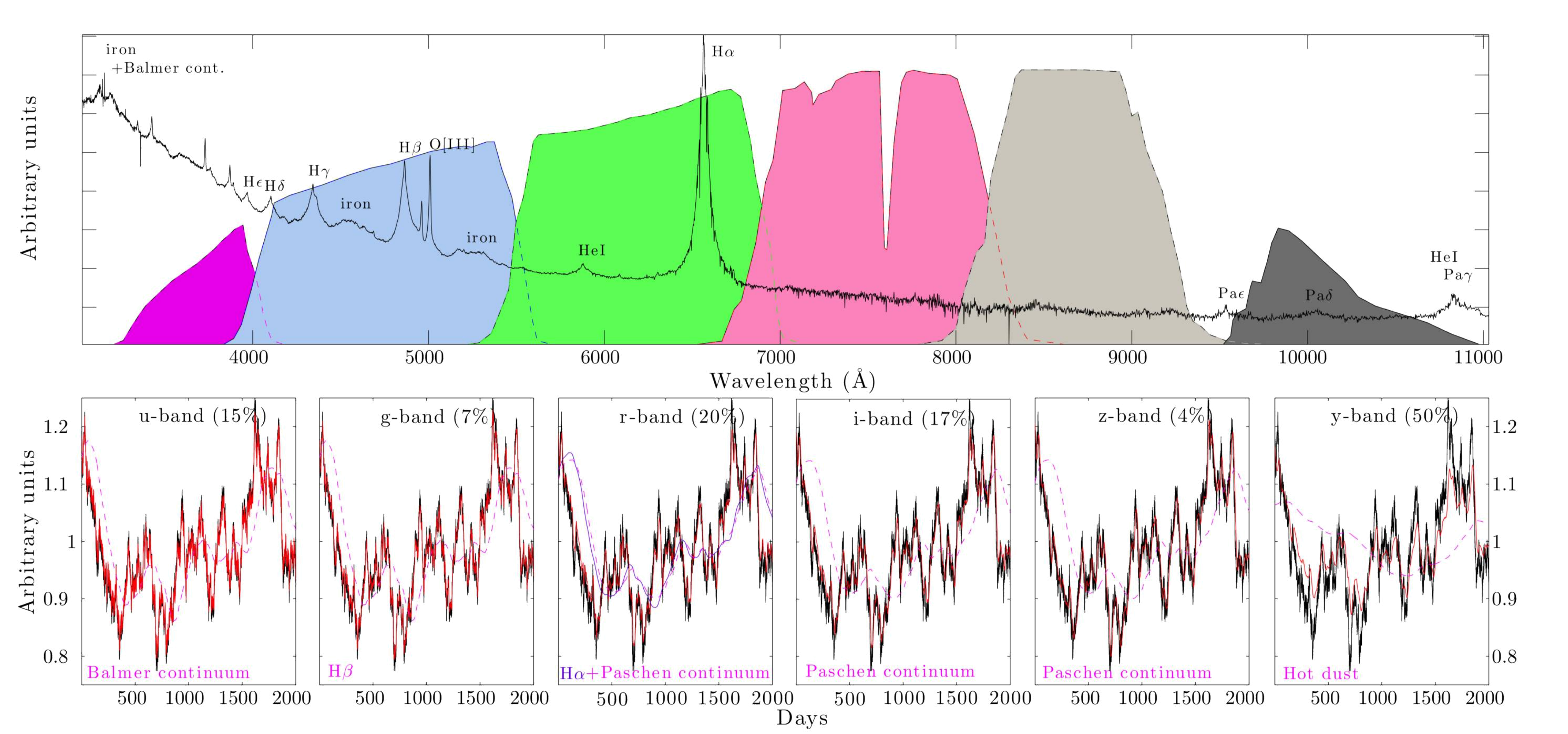}
\caption{Light curve simulations in photometric bands. Upper panel shows a typical $z=0$ AGN spectrum (\citealt{dvb01,gl06}) and the contribution of the various emission features to the broadband flux assuming LSST throughput curves. Bottom panels exhibit the broadband light curves in individual panels (red curves) with the total contribution of BLR emission features denoted in per cent in each panel. For clarity, measurement uncertainties are not shown, but are included in the final mock light-curve products (see Fig. \ref{lcs}). For comparison, the driving continuum light curve is shown in black line in each panel, with the normalized light curves of prominent emission features in dashed magenta lines (and purple solid line for the H$\alpha$ light curve contributing to the $r$-band). Note the smoother appearance of the light curve in the redder bands due to signal transfer effects in a geometrically-thick region (the effect of reduced variability with increasing wavelength is separately and phenomenologically treated in \S3).}
\label{fil}
\end{figure*}

Once the wavelength-dependent (ionizing) continuum emission from the accretion disk has been defined for all bands, the contribution of emission lines and line blends, as well as variable non-ionizing continuum emission from the outer regions (e.g., Balmer and Paschen continuum emission from the BLR, \citealt{kor01}, and heated dust emission from its far outskirts, \citealt{nl93}) must be included. To this end, we compiled a list of major emission components, and estimated their relative contribution and time delays with respect to the adjacent continuum level from a mean quasar composite spectrum \citep{dvb01}, and from various reverberation mapping campaigns \citep{kas00,pet04,met06,sug06,kas07,ben10,bar13,cr13}\footnote{We resort to educated guesses and theoretical works - e.g., \citet{nl93} - to estimate the lags in lines and blends for which reverberation results are currently unavailable.}. We note that we consider luminosity-independent rest equivalent width values in this work, which is a clear simplification given the scatter observed \citep{sh11} and the presence of the Baldwin effect, i.e., the anti correlation between line equivalent width and AGN luminosity \citep{bal77,bia12}. As our emission line properties are deduced from SDSS data for relatively bright objects \citep{dvb01}, we expect the median equivalent width for LSST objects to be comparable to, or somewhat larger than, that assumed here\footnote{It turns out that the luminosity of the bulk of the quasars for which reverberation mapping would be feasible with LSST is $\gtrsim 10^{45}\,{\rm erg~s^{-1}}$, hence consistent with the SDSS composite \citep{dvb01}.}. Throughout this work, we assume that the time delays for a given emission line (or blend) scale as $L_{\rm opt}^{0.5}$ \citep[but see \citealt{kas05}]{ben09,cr13}, and that the variance in the line light curves is comparable to that in the continuum \citep{kas00,woo08}. Table 1 lists the various emission components used in this work, to which we refer to henceforth as emission lines.

The total AGN light curve in band $j$ is then given by: 
\begin{equation}
\displaystyle f^j(t)=f_c^j(t)+\sum_{i=1}^N\left [ f_c ^j* \psi_i(\tau_i(L,z)) \right ] \frac{\bar{\eta}_j(\lambda_i(z))W_0^i(z)}{\int d\lambda \eta_j(\lambda)}
\end{equation}
where the summation is over all entries in Table 1. $\psi_i$ is the line transfer function for the $i$'th emission feature, and is assumed to be of a rectangular form (see above), which starts at zero lag, and extends out to $2\tau_i$ ($\tau$ is the time delay). The telescope throughput curve for band $j$, $\eta_j(\lambda)$, is shown in Fig. \ref{fil} for the proposed set of LSST filters (including telescope efficiency and atmospheric effects).  The rest equivalent width values for the relevant transitions, $W_0$, are given in Table 1. We define the mean efficiency for an emission component centered on $\lambda_i$ as,
\begin{equation}
\bar{\eta}_j(\lambda_i)=\frac{\int d\lambda \eta_j(\lambda) \phi(\lambda-\lambda_i)}{\int d\lambda \phi(\lambda-\lambda_i)}
\end{equation}
where $\phi(\lambda)$ is the profile of the emission component. For simplicity, we assume rectangular profiles with a width $\delta \lambda=30$\AA\ for the emission lines (this corresponds to velocities of $\gtrsim 10^3\,{\rm km~s^{-1}}$, typical of broad line widths). For large scale continuum emission (e.g., the Balmer bump), we  assume a rectangular profile which spans the wavelength range given in Table 1 (rows 32--34).

Examples for the light curves in different LSST bands for a $z=0,~L_{\rm opt,45}=1$ AGN are shown in Figure \ref{fil}. For clarity, we do not consider in this example the decreasing variance with wavelength (Eq. \ref{ep}) and assume measurement uncertainties are negligible (those effects {\it are} included in the full calculations presented in sections 3, 4). Clearly, light curves in all bands are, to zeroth order, similar due to the dominant contribution of correlated continuum processes. Nevertheless, finite differences are noticeable: at progressively redder bands, the light curves are less spiky due to transfer effects across the accretion disk, which suppress small scale fluctuations (e.g., compare the light curve in the $y$ band and the driving continuum signal). In addition, the finite contribution of emission lines and blends to the broadband flux leads to longer trend deviations, and further suppresses small scale fluctuations. For example, the (normalized) $r$-band light curve lies above the driving continuum level at $<500$\,days, but below the driving continuum light curve at $\sim 1000$\,days, which is due to the lagging contribution of the H$\alpha$ line to the flux. A more prominent effect is seen in the $y$ band, where a considerably delayed large-amplitude contribution from dust emission is apparent. Such differences between light curves in different bands can be detected in good quality data, allowing for the measurement of time delays associated with major emission lines \citep{cd11,cz13}.

\begin{figure}
\epsscale{1.2}
\plotone{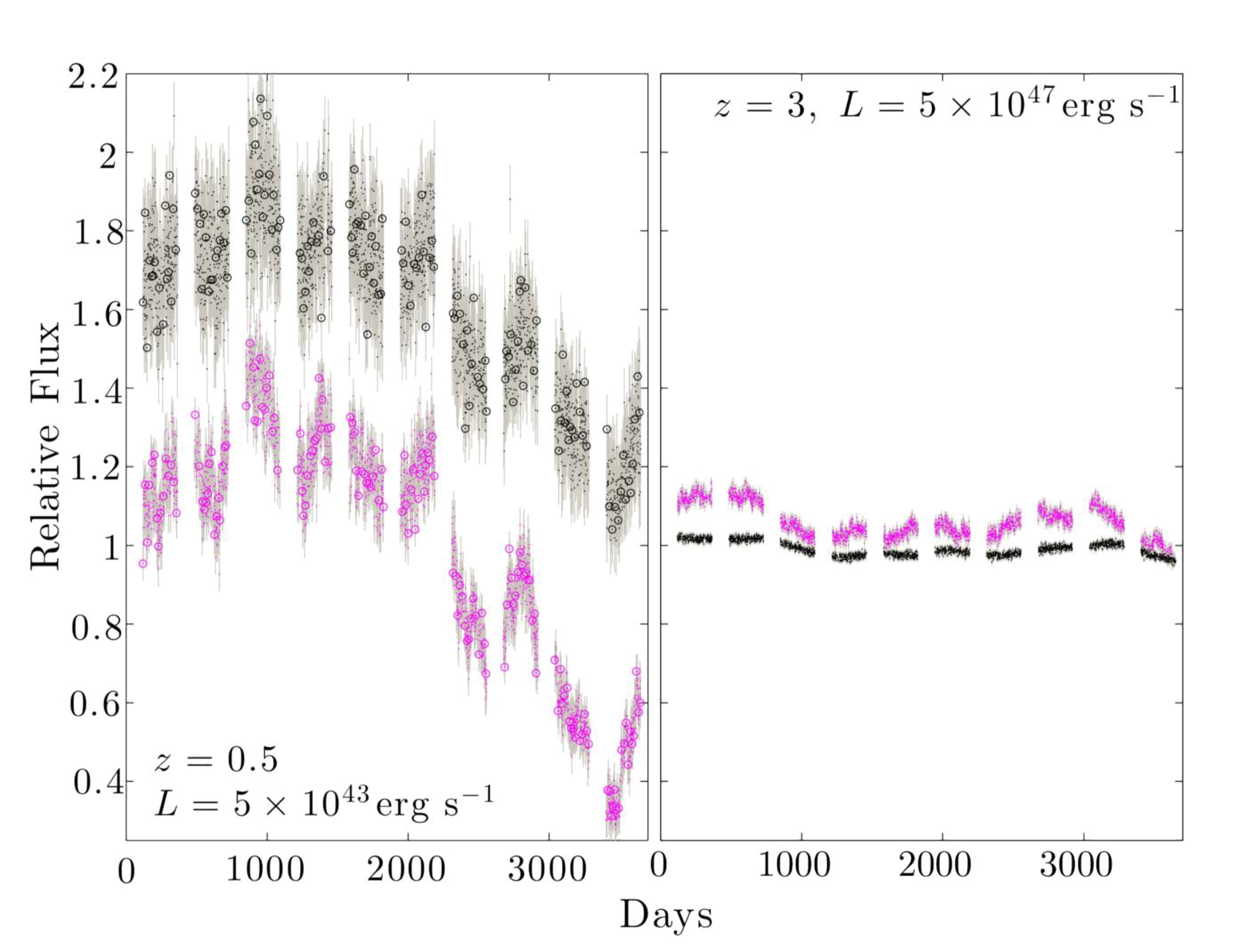}
\caption{Representative mock light curves for the $u$ band (magenta) and the $y$ band (black) for a $z=0.5$ Seyfert galaxy (left column) and a bright quasar at $z=3$ (arbitrary shifts between the broadband light curves were introduced for clarity) using 15\,sec exposures with LSST. DDF light curves are shown as dots while UNS as circles (only for the Seyfert galaxy case). Note the markedly different variability amplitude of luminous vs. less luminous objects, and the generally higher variance at shorter wavelengths (see text).}
\label{lcs}
\end{figure}

\subsection{Sampling and Measurement Noise}

With LSST being the largest planned photometric survey in the coming decades, we consider the following observing patterns, which are inspired by the current LSST operations simulator (Opsim\footnote{\url{https://www.lsstcorp.org/opsim/home}}): a) deep drilling fields (DDF) cadence, and b) universal sampling (UNS). For the DDF, daily contemporaneous acquisition of data in all bands is considered, but with two variants: with 120\,days seasonal gaps (characterizing a fair fraction of the observable sky), or in a continuous viewing mode (e.g., for fields aimed toward the pole). For the UNS, we take the representative number of visits in each band over the sky, as given by the current Opsim, so that $u/g/r/i/z/y$ bands have $56/80/184/184/160/160$ visits over the entire ten-year duration of the survey. We note, however, that these are mere averages, and that some directions over the sky will likely benefit from better sampling while others will be characterized by sparser light curves. Four months seasonal gaps are assumed for all UNS sources but otherwise regular sampling is considered per band. We do not simulate data gaps due to downtime of the telescope, nor do we consider more realistic sampling patterns for the LSST, as those are, currently, publicly unavailable.

To estimate the noise level of a given observation, in each filter, we follow the results of the LSST exposure time calculator (ETC\footnote{\url{http://dls.physics.ucdavis.edu/etc/}}), which provides approximate (average) values for the  background noise, dark current noise, and readout noise for a 15\,sec on (point) source exposure. The effect of varying background resulting from object declination and time of observation, as well as weather conditions, is not considered here. We assume Poisson noise associated with the target signal, and normalize according to the ETC for each band. The total measurement noise of an observation is obtained by summing all sources of noise in quadrature. 

Representative lightcurves for a $z=0.5$ Seyfert 1 galaxy, and a $z=3$ quasar in the $u$ and $y$ bands are shown in figure \ref{lcs} (we do not include the effect of intervening systems, such as the Ly$\alpha$ forest, on the flux level of high-$z$ sources).  In the examples shown, the light curve of the low-$z$ Seyfert galaxy is considerably noisier than the quasar's, with the latter being characterized by a smaller variance. The effect of seasonal gaps is also evident.

\subsubsection{Host Contribution}

Finite aperture observations result in some of the host light contributing to the signal, and may lead to additional  noise under varying seeing conditions\footnote{Note that a constant host contribution to the light curves is immaterial as reverberation mapping algorithms are insensitive to it \citep{wel99,cz13}.}. For high-$z$ sources, being the bulk of the AGN population in upcoming surveys (see Section 4), the contribution of the host decreases sharply because the surface brightness scales as $(1+z)^{-4}$ and galaxies are rest-UV faint
relative to the nuclear emission. To mitigate host contamination problems in faint low-$z$ objects, large-aperture photometry or image-subtraction techniques \citep{al98} provide viable solutions. For the LSST, image-subtraction techniques will be part of the standard reduction pipeline, and are expected to reduce varying seeing effects down to photon noise levels.  Simulating second-order host galaxy subtraction effects, which could affect  a small fraction of the quasar population, is beyond the scope of the present work and will not alter the main conclusions of this paper.  

\subsection{Analysis}

We apply the photometric reverberation mapping technique of \citet{cz13} to our mock photometric light curves. Briefly, one considers a pair of filters, one of which is relatively line rich, while the other one is line poor, and calculates the multi-variate correlation function (MCF) to deduce a lag. In cases where the non-continuum emission contribution to the line-rich band is dominated by a single emission line, while the other (line-poor) band is devoid of emission components other than the primary ionizing continuum, the lag is the line-to-continuum time delay to a good approximation \citep{cz13}. Denoting the line-poor (fluxed) light curve as $f_c(t)$, and the line-rich (fluxed) light curve as $f_{cl}(t)$, then, to first-order approximation, the model for $f_{cl}(t)$ is
\begin{equation}
f_{cl}^m(t)=(1-\alpha)f_c(t)+\alpha f_c(t-\tau),
\label{fclm}
\end{equation}
where $\alpha$ is the relative emission line contribution to the signal in the band, and $\tau$ is the line-to-continuum lag. One may deduce both $\alpha$ and $\tau$ by searching for the peak of the MCF defined by 
\begin{equation}
R(\alpha,\tau)=\frac{1}{N\sigma(f_{cl})\sigma(f_{cl}^m)}\sum_i^N f_{cl}(t_i)f_{cl}^m(t_i,\tau,\alpha),
\label{R}
\end{equation}
where $\sigma$ is the standard deviation of the time series, and $N$ is the number of points in the sum  \citep[see][and references therein for further details]{cz13}. Unlike \citet{cz13}, we do not carry out Monte Carlo simulations on a case-by-case basis to test for the significance of individual solutions since these would result in prohibitively-long execution times on current hardware. As we shall show below (section 3.1), it is possible to effectively screen against unphysical solutions\footnote{The degree to which improved lag statistics for large samples of AGN may be obtained by considering the significance of individual measurements  is worth exploring in future studies. This awaits better error-estimation algorithms for reverberation mapping, and further improvements in computer hardware.}. 

While the MCF method has been shown to work well in regions of the spectrum with relatively sparse emission line density (e.g., for low-redshift objects in the redder bands), it is yet to be established whether it can also be applied at higher-$z$, where the  optical wavelength bands are populated by many, possibly overlapping emission lines and blends, each of which is characterized by its own time delay and relative contribution to the flux. In such instances the aforementioned formalism is not strictly valid, yet could still serve as a fair approximation to the ideal experiment. The degree to which this statement applies is quantitatively explored in this work. 

It also needs to be demonstrated that the MCF formalism of \citet{cz13} can work under more realistic observing conditions, which may be characterized by sparse, non-contemporaneous  sampling in different bands. Further, redshift determination by photometric means alone has finite accuracy \citep[and note that the expected photometric redshift accuracy for LSST quasars is at the $\sim 5$\% level hence of the order of the typical emission line width\footnote{\url{http://www.lsst.org/files/docs/sciencebook/SB\_10.pdf}}]{r01,br13}, and so the choice of line-rich and line-poor bands may be erroneous for certain redshift intervals, and for a fraction of all objects. The degree to which such issues affect line-to-continuum time-delay measurements is explored below.

\subsubsection{Implementation}

\begin{figure}
\epsscale{1.2}
\plotone{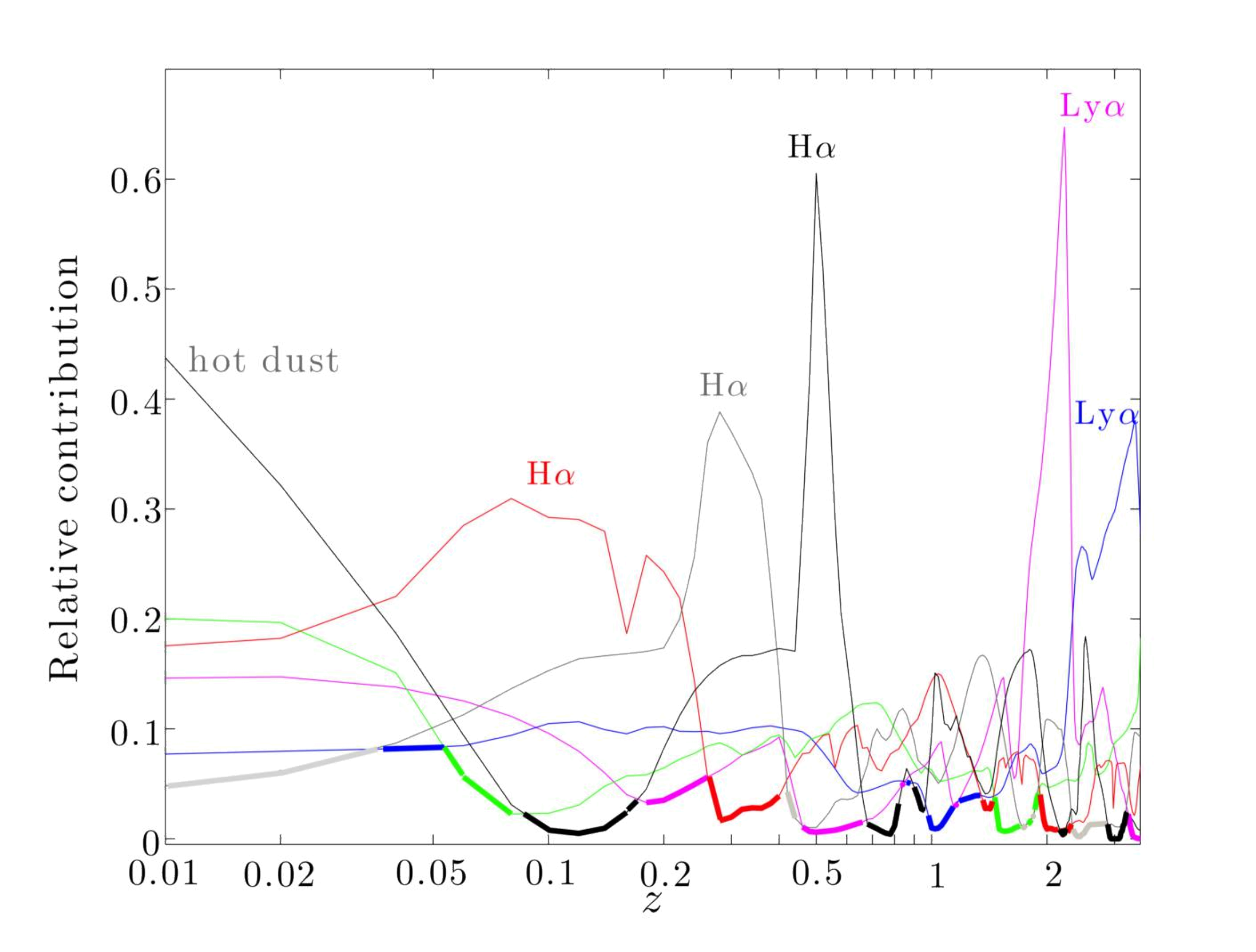}
\caption{The relative contribution of BLR emission components (i.e., spectral lines and continua features) to each of the LSST bands (color coding is as in Fig. \ref{fil}). The peaks correspond to redshift intervals where either the H$\alpha$ or the Ly$\alpha$ contributions to the bands dominate. Thickened curves mark the redshift interval over which bands may be considered as relatively line-poor bands (i.e., as $f_{c}$ in Eq. \ref{fclm}; see text). For $z>2$ objects, the Ly$\alpha$ forest becomes increasingly important, but its effect is not included. Similarly, the contribution of (non-varying) host emission to the flux in the bands is not taken into account (see text).}
\label{frac}
\end{figure}

Given light curves in six filters per object, we first identify line-poor and line-rich bands so that the above model (Eq.  \ref{fclm}) is justified and the analysis meaningful. 

Typical emission line contribution to the bands is of order 10\% \citep{cd11}, but can reach 30\% for $z<0.5$ sources where the H$\alpha$ and the Paschen bump contribution are dominant (Fig. \ref{frac}). Line emission may even dominate the flux in the (relatively narrow) $u$ and $y$ bands at redshift intervals where the strongest hydrogen emission lines, Ly$\alpha$ and H$\alpha$, are relevant.

While no band may be strictly free from BLR emission, we identify $f_c$ with the band characterized by the smallest contribution of BLR emission features to its flux (e.g., the $y$-band for $0.1<z<0.2$ sources using our emission model; see Fig. \ref{frac}). It should be emphasized that the suitability of any given band to serve as $f_c$ depends not only on the relative emission line contribution to the flux, but also on the host galaxy contribution to it and the effect of seeing in nearby faint AGN, as well as on the number of visits, and how far in wavelength the band is removed from the line-rich band in question (see more \S3.4). Once $f_c$ is defined, time-delay measurements proceed via the calculation of the MCF with respect to each of the line-richer bands, resulting in up to five time-delay measurements per object. 

In this work we take the following approach to benchmark photometric reverberation mapping: the deduced time delay is compared to the (input) time delay of the emission component which contributes most to the line-rich band. This is certainly a reasonable interpretation when one dominant emission line exists, and the degree to which it is generally useful is critically examined below (section 3.3).

\section{Results}

\begin{figure}
\epsscale{1.2}
\plotone{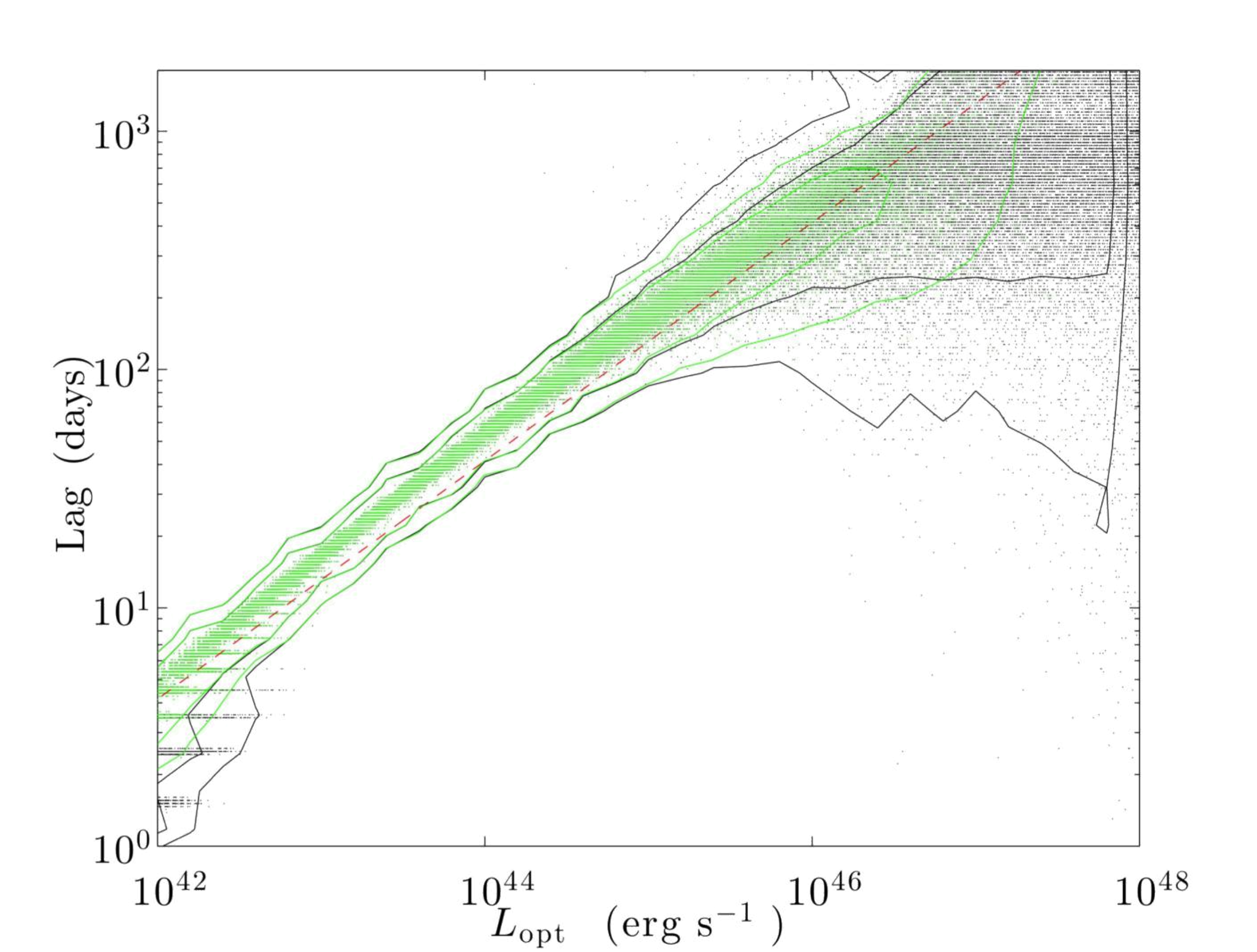}
\caption{The H$\alpha$ radius-luminosity relation for $z=0.01$ objects. The input relation is shown in dashed red line, and individual measurements for $10^5$ AGN using the $z$ filter as the line-poor band, and the $r$ filter as the line-rich band, with DDF sampling, are shown in black points. Discretization effects are evident at short lags, while the undersampling of the full extent of the transfer function is seen at the high-luminosity end. When filtering for objects whose solution for the relative emission line contribution to the band, $\alpha$, is within 40\% of the input value, outliers are naturally discarded, resulting in statistically more accurate  lag solutions (green points). There is an overall shift with respect to the input relation as is discussed in the text. }
\label{OE_RL}
\end{figure}

\begin{figure}
\epsscale{1.2}
\plotone{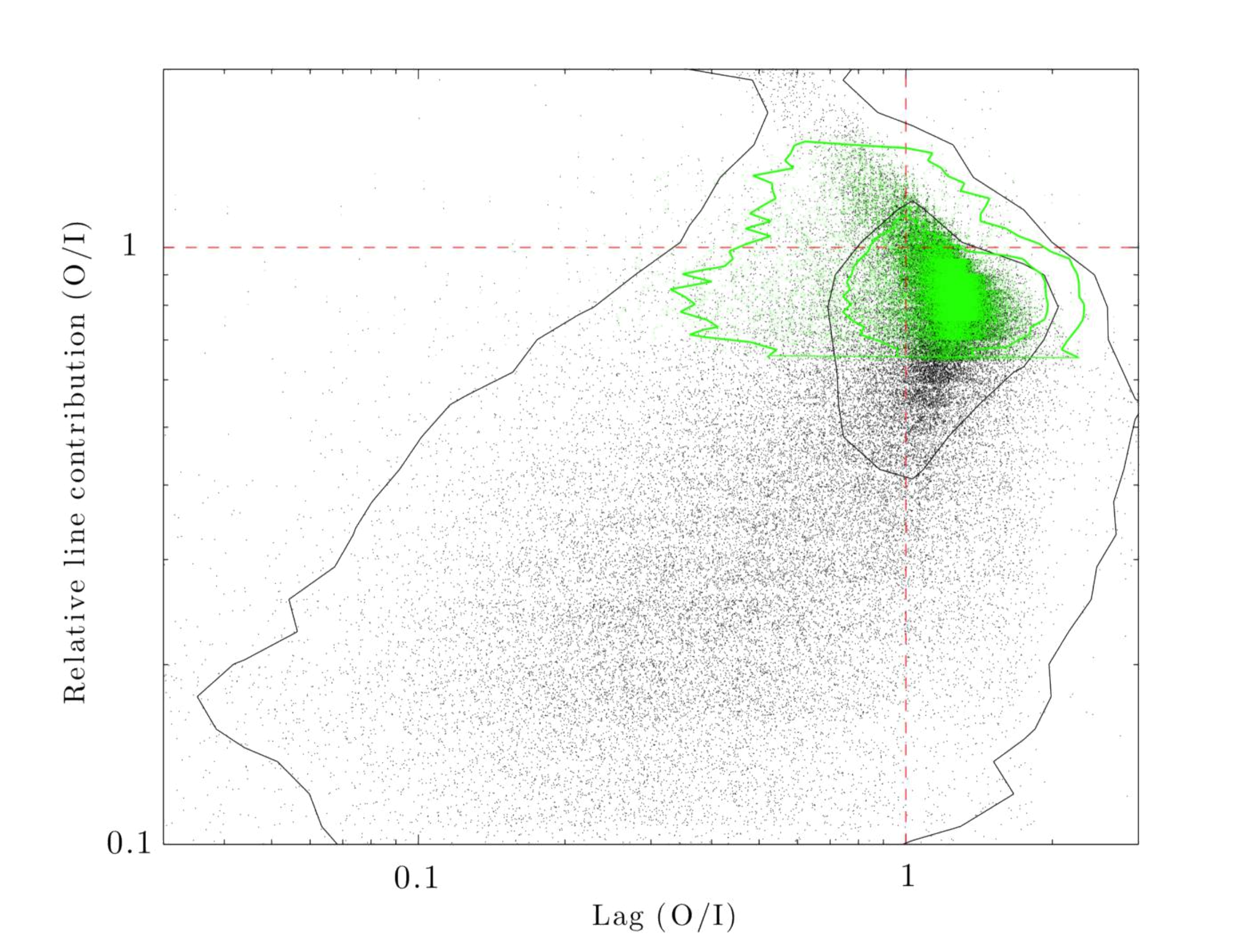}
\caption{The output (i.e., recovered) to input (O/I) ratios for the H$\alpha$ line properties for all measurements (black points) and those for which the recovered relative line contribution to the flux in the band, $\alpha$, deviates by less than 40\% from the input value (green points). Clearly, filtering against bad $\alpha$ solutions considerably reduces the scatter in the lag measurements. Specifically, most solutions fall around O/I$\sim 1$ in either of the quantities, yet a shift, of order 20\% is present, which results in an overestimation of the lag, and an underestimation of $\alpha$ (see text).}
\label{OE_Q}
\end{figure}

\subsection{Single Emission Line}

We first consider the simplest case, where a single emission line, H$\alpha$, is present in the spectrum, and attempt to recover its delay from photometric light curves that are characterized by DDF sampling in continuous viewing mode. For simplicity, all quasars are placed at $z=0.01$ (i.e., resulting in high signal-to-noise light curves), and span a broad range of luminosity\footnote{We do not consider luminosity-function statistics in this section (see \S4).}; in this case, $f_{cl}$ is the $r$-band light curve, and we arbitrarily assign the $z$-band light curve to $f_c$. Unless otherwise stated, of order $10^5$ objects were simulated, and the MCF was calculated for each, resulting in the time-delay statistics provided below.

Figure \ref{OE_RL} shows the recovered radius-luminosity relation\footnote{Realistically, the BLR sizes for a given emission line in quasars of a given set of properties, such as luminosity, follow a physical distribution of some form rather than reflect on a single value. The resulting distribution functions for the recovered lags could therefore be somewhat broader than calculated here. Currently higher moments of the BLR size-luminosity relations are poorly understood hence not included in our model.} for H$\alpha$. On average, the slope of the lag-luminosity relation is well traced over the intermediate luminosity range. In particular, the deduced slope of the size-luminosity relation for $\sim 30,000$ objects with $10^{43}\,{\rm erg~s^{-1}}<L_{\rm opt}<10^{45}\,{\rm erg~s^{-1}}$ is $0.502\pm0.003$, which is in excellent agreement with the input value\footnote{The uncertainty was determined by considering sub samples of the data, and obtaining fit statistics by means of bootstrapping.}. At the low-luminosity end, the effects of sampling discretization are evident as time-lags approach the sampling period, and the transfer function is under sampled. At the high-luminosity end, clear deviations from the input size-luminosity relation are evident as the survey's lifetime (10 years for the first phase of the LSST) is comparable to the lag, and the full extent of the transfer function cannot be observed. In particular, the recovered lag distribution for $L>10^{46},{\rm erg~s^{-1}}$ objects is considerably skewed to unphysically short lags. As we discuss below, reliable time-delay measurements are possible provided additional information exists.

Considering the ratio between the output (i.e., recovered) and input quantities (O/I) for quasars of all luminosities, we see that most points cluster, as expected, around ${\rm O/I}\sim1$ (Fig. \ref{OE_Q}). In particular, lag measurements that considerably deviate from the input lag are very likely to result in an offset solution for $\alpha$: e.g., lag solutions for the most luminous sources in figure \ref{OE_RL} are skewed to shorter lags and to low $\alpha$ values, which do not match the input values. Therefore, provided independent information on $\alpha$ is available - for example, through spectroscopic observations, or from statistical knowledge of the quasar population as a whole - it is possible to screen against non-physical solutions by rejecting highly discrepant $\alpha$ values. Generally, allowing for smaller uncertainty margins on $\alpha$ results in a smaller scatter in the lag O/I statistics, and also with fewer objects contributing to the sample (not shown). 

As "valid solutions" in this work, we consider $\alpha$ values that deviate from the input value by no more than $\pm40\%$. This criterion seems to provide satisfactory results for a large range of object redshifts and luminosities (see section 4), and is consistent with the results of \citet{sh11} who show that $\sim 90$\% of all H$\alpha$ rest equivalent width measurements for SDSS quasars lie within a similar interval. Valid solutions under our $\alpha$-interval definition are relatively well clustered around the input values as can be seen in Figure \ref{OE_RL}. In particular, even for the most luminous sources, whose line transfer function is poorly sampled by the data, it is possible to identify reliable solutions in a fraction of all objects (note how the green points follow the input relation up to high luminosities in figure \ref{OE_RL}).

\begin{figure}
\epsscale{1.2}
\plotone{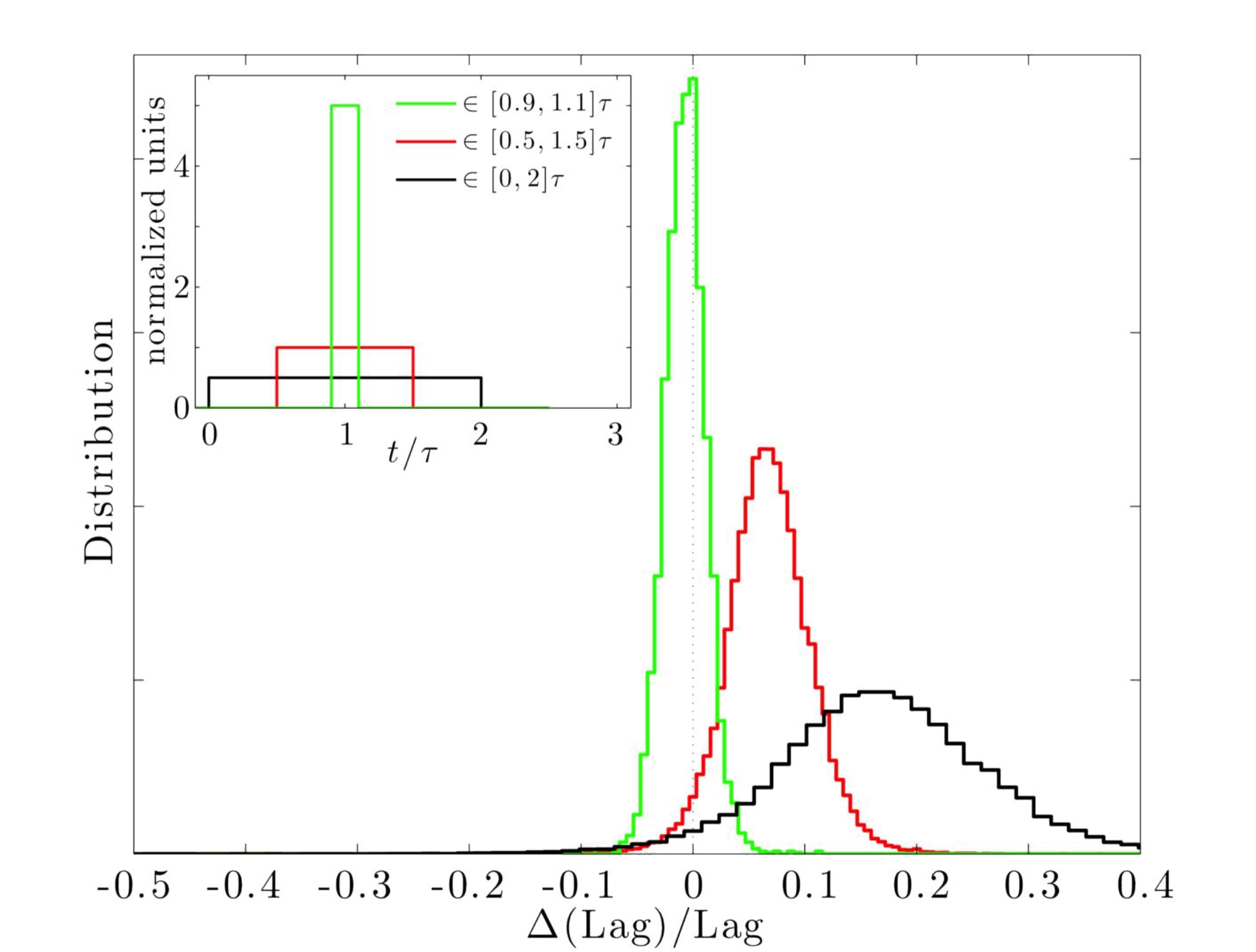}
\caption{The bias in lag determination. Different colors correspond to the different lag distributions obtained under different assumptions regarding the shape of  the transfer functions (see inset). Transfer functions which are farther removed from zero lags, corresponding to less BLR material along the sightline to the observer, result in a smaller lag bias (see text). Changing the power density spectrum of the object does not substantially alter these results (Figure \ref{pow}).}
\label{trans}
\end{figure}

\begin{figure}
\epsscale{1.2}
\plotone{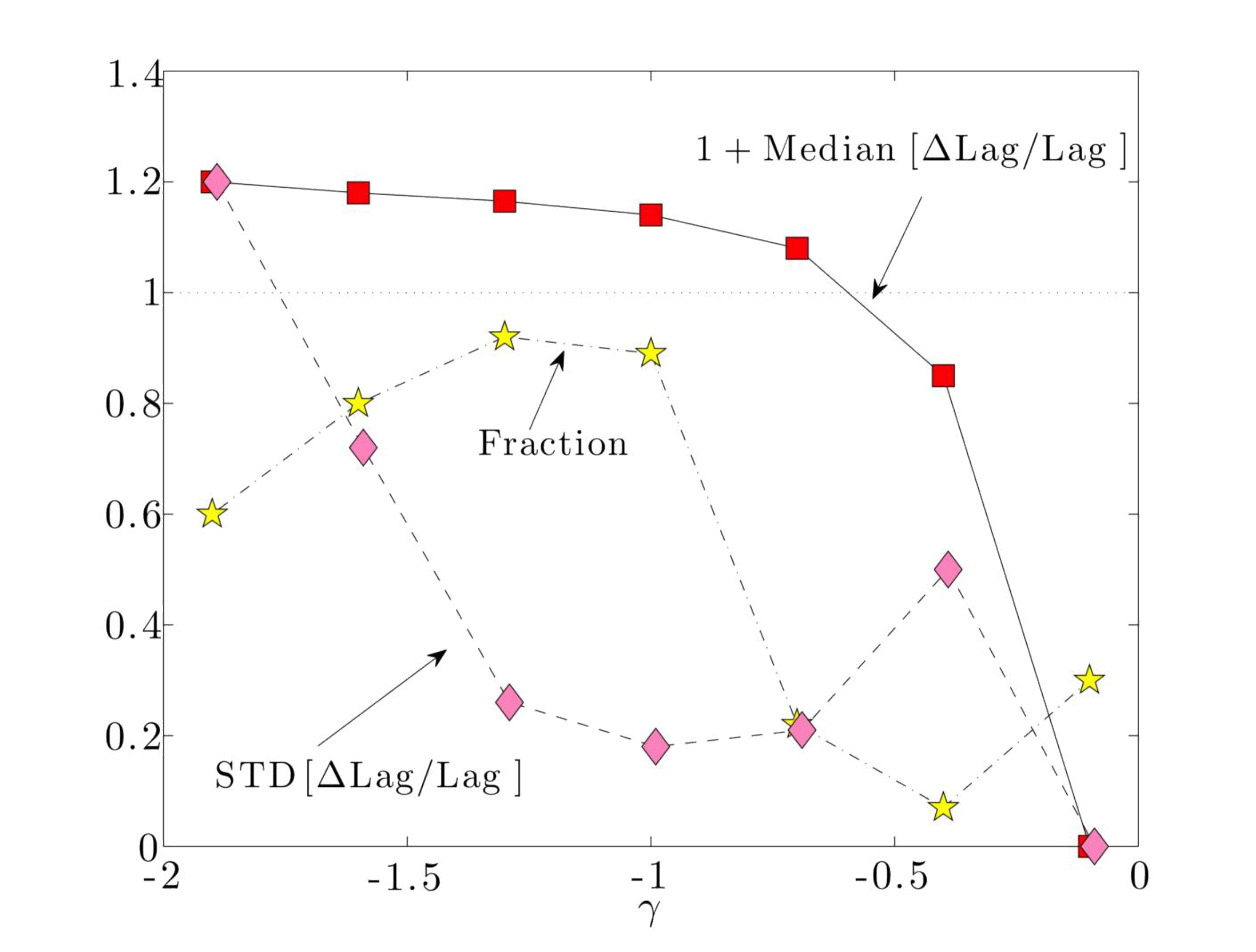}
\caption{The dependence of the recovered-lag statistics on the power density spectrum. For $-1.3<\gamma<-0.8$ the lag distribution is narrow, and the fraction of objects with a reliable lag determination peaks. The ability of the algorithm to deduce the lag is significantly reduced for $\gamma >-0.4$, as light curves resemble white noise. For AGN with the reddest power spectra, individual measurements are less reliable, yet the median traces the input lag. In this limit, the lag distribution is non Gaussian.}
\label{pow}
\end{figure}

There is a statistically significant, and consistent, $\lesssim 20\%$ overestimation of the median lag with $\alpha$ being underestimated by a similar factor (see the bulk of points in Fig. \ref{OE_Q}). This small but significant effect can only be detected with good enough statistics and reliable independent lag measurements, hence could not have been determined by \citet{cz13} in their analysis of \citet{ser05} data. The underlying reason for these offsets has to do with the MCF formalism being effectively blind to the structure of the line transfer function on short timescales. In particular, any emission line contribution to the light curve at short times (e.g., due to clouds which lie closer to our sightline and whose reprocessed radiation arrives nearly simultaneously with the continuum light curve), may be regarded as pure continuum emission. This is not surprising since line and continuum emission cannot be independently traced by photometric data. Considering narrower emission-line transfer functions, whose centroids are identical yet rise at finite times - i.e., the reverberating signal is sufficiently  removed from zero lags - results in a reduced bias (Fig. \ref{trans}; see also \citealt{zu13} who use a narrow Gaussian transfer function). Similarly, the small offset in $\alpha$ also vanishes, indicating its common origin; by not being able to trace part of the transfer function, the relative contribution of the line to the photometric light curve is effectively diminished. This in turn means that some information about the transfer function of emission lines is lurking also in broadband photometric data.

\begin{figure*}
\epsscale{1.22}
\plotone{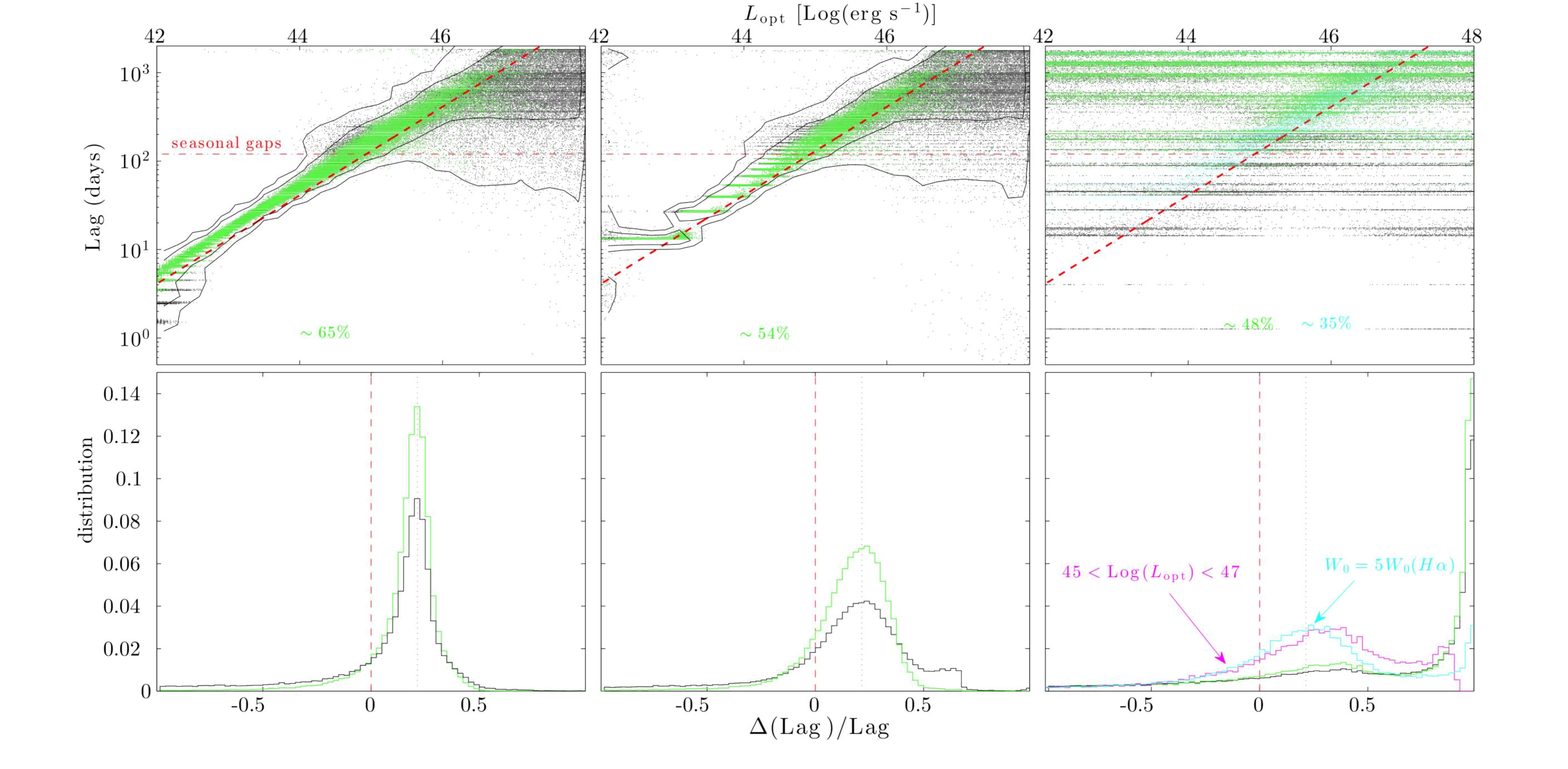}
\caption{$R_{\rm BLR}(L)$ determination for $z=0.01$ objects for several sampling patterns (upper panels), and recovered time-lag distributions with respect to the input lag (lower panels). Left column corresponds to DDF sampling with 120\,days gaps (c.f. Figure \ref{OE_RL}). The middle column is akin to UNS fields where we assume 200\,visits per band, with seasonal gaps. Right panels corresponds to UNS fields with 60\,visits in one band, and 200\,visits in the other band over the LSST lifetime (the identity of the filters is unimportant, and contours are not shown, for clarity). Clearly, as sampling becomes sparser, the scatter is increased, and the time-delay distributions around the input value become broader. It is possible, however, to narrow the distributions by rejecting non-physical solutions based on prior knowledge of $\alpha$ (here we require that the recovered $\alpha$ is within $\pm40$\% of the input value). Such solutions and their respective distributions are shown as green points and curves in all panels, and the fraction of good solutions is denoted in percent in each of the upper panels. For the right panels we also show solutions and distributions focusing on intermediate luminosity objects, which do not suffer as much from finite sampling issues (magenta curves). Further, we consider a case in which the emission line is assumed to have a larger contribution to the band yet with all other parameters being fixed (cyan points and curves).}
\label{lll}
\end{figure*}

Lastly, we wish to study the effect of the power-density spectral shape, $\gamma$, on the ability of the MCF algorithm to reliably recover time delays. In particular, a considerable range of powerlaw indices have been found to characterize quasar lightcurves, with slopes, $-1.7<\gamma<-0.4$ \citep{giv99,mu11}.  To this end we consider a narrow range of intermediate luminosity objects (so that sampling issues are irrelevant), and plot the statistics of the recovered signal in Figure \ref{pow}. As shown, the $\gamma$ range covered by most AGN lies around a sweet spot in the parameter space, where the scatter in individual time delays around the median value is minimal, and the fraction of objects for which reliable time-lag solutions are obtained peaks (as before, we use the constraint that the recovered $\alpha$ deviates by less than $\pm40$\% from the input value). For objects whose light curves show considerable variability only at the longest timescales (i.e., $\gamma<-1.6$), the recovered lags show the greatest dispersion around the median lag: while the latter quantity is robust, individual lag measurements are less reliable. Similarly, the fraction of unphysical solutions is somewhat increased for small $\gamma$ values. At the other extreme, where AGN light curves gradually resemble white noise ($\gamma>-0.5$) , the fraction of objects with physical solutions drops significantly, and the dispersion in the lag increases so that lag measurements become unreliable. Moreover, even statistical averages become meaningless as the algorithm prefers to fit white noise with white noise, and the median time delay goes to zero. The slight increase in the fraction of "reliable" physical solutions that satisfy our $\alpha$ criterion for $\gamma >-0.4$ (note the last bin in Fig. \ref{pow}) is purely artificial and has to do with the fact that the recovered $\alpha$ in this limit is drawn from  a uniform distribution in the range $[0,1]$. We note that the aforementioned (transfer-function dependent) offset in the median lag depends relatively little on $\gamma$ for the parameter range typical of quasars \citep{giv99,mu11}.

\subsection{Sampling}

Realistic experiments rarely benefit from strictly uniform sampling since seasonal gaps, telescope down times, bad weather, and scheduling constraints  are all limiting factors. The effect of different sampling patterns on the size-luminosity relation for H$\alpha$ (again, neglecting other lines), are shown in Figure \ref{lll}. Generally, with more timescales characterizing the observing pattern, the noisier the measurement becomes. In particular, introducing 120\,day seasonal gaps results in noisier lag measurements around those timescales, which act as effective solution attractors. Interestingly, there is a tendency for the algorithm to somewhat overestimate the lag around the gap timescale (Fig. \ref{lll}). Again, erroneous lag measurements can be discarded using additional information, if available, on $\alpha$: screening against discrepant $\alpha$ solutions, some $\sim 35$\% of all measurements are discarded, and the remaining solutions match well the input size-luminosity relation of the BLR. Specifically, the cluster of overestimated time-delay solutions around seasonal gap timescales is naturally excluded by this filtering scheme. 

Experimenting instead with UNS-like sampling for the LSST, and restricting our analysis to light curves with $200$ visits in each band, we find that the ability to recover short delays (e.g., in faint enough AGN or in high-ionization emission lines) is compromised. In particular, the sampling period in this case is $\sim 20$\,days,  which is of the order of the minimal lag that can be reliably recovered. As before, using prior knowledge on $\alpha$ reduces the scatter, especially at the low-luminosity end of the quasar population, which is characterized by the shortest lags, and allows for a better characterization of the size-luminosity relation.

\begin{figure*}
\epsscale{1.22}
\plotone{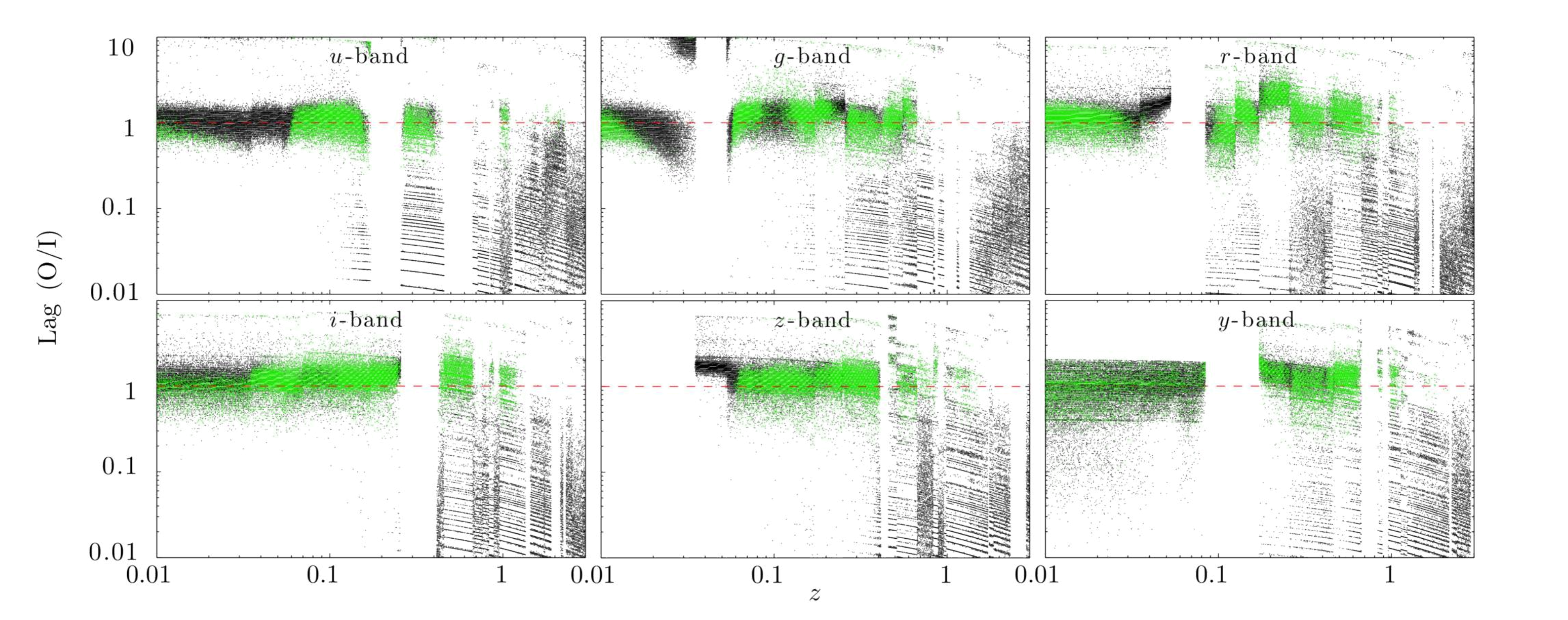}
\caption{Lag determination, in relative units (O/I), as a function of redshift, for different bands, and for quasars with $L_{\rm opt}=10^{45.5}\,{\rm erg~s^{-1}}$. DDF sampling with 15\,sec exposures and 120\,days seasonal gaps are assumed. Black points mark all solutions, while green points constitute a sub-sample which agrees with our criterion for adequate $\alpha$ solutions (see text). Clearly, there are broad redshift intervals where the recovered lag is in good agreement with the input lag up to a modest bias (see text). For a specific band, redshift gaps having no solutions mark intervals where the filter is used as the line-free band. The sharp limit on O/I for the $y$ band, at low $z$, results from an enforced time-delay cutoff in the phase space probed by the MCF, which corresponds to half of the LSST lifetime, and the dust component, which contributes to this band at those redshifts having $\sim 500$\,days delay. The ability of the algorithm to recover the lag at high-redshifts is reduced by low S/N and the finite lifetime of the survey (a small cluster of green points for the $u$-band around $z\simeq 2$ corresponds to Ly$\alpha$ lag measurements, which are limited by S/N given the source luminosity; see also \S4).}
\label{zzz}
\end{figure*}

Lastly, we consider UNS, with one of the bands having $200$\,visits, and the other band a mere $60$\,visits\footnote{We do not find significant differences between cases in which the sampling is interchanged between the line-rich and line-poor bands.} (e.g., the $r$ and $u$ filters given the current version of the Opsim). In this case, the scatter is considerably increased (note the highly broadened lag distributions in Figure \ref{lll}), as is the minimal time delay that can be reliably recovered; e.g., measuring the H$\alpha$ lag for low-luminosity AGN is impossible. Discarding non-physical solutions using our current screening algorithm is effective over a relatively narrow luminosity range ($10^{45}\,{\rm erg~s^{-1}}<L<10^{46}\,{\rm erg~s^{-1}}$) but even then individual measurements contain little information, with only large ensembles being able to characterize the typical time delay up to a $\sim 40$\% bias. Our calculations show that further screening the sample for objects showing the greatest variance in their light curves may be useful to further narrow down the lag distributions yet it is not clear that such a selection would provide an unbiased view of the quasar population as a whole, hence it is not considered here. Additional tests for the MCF solution may prove beneficial \citep[see][for one possible version of such a test]{cz13} in further weeding out erroneous solutions. These, however, are beyond the scope of the present work due to their prohibitive computational demands (see also section 2.4). 

\begin{figure*}
\epsscale{1.22}
\plotone{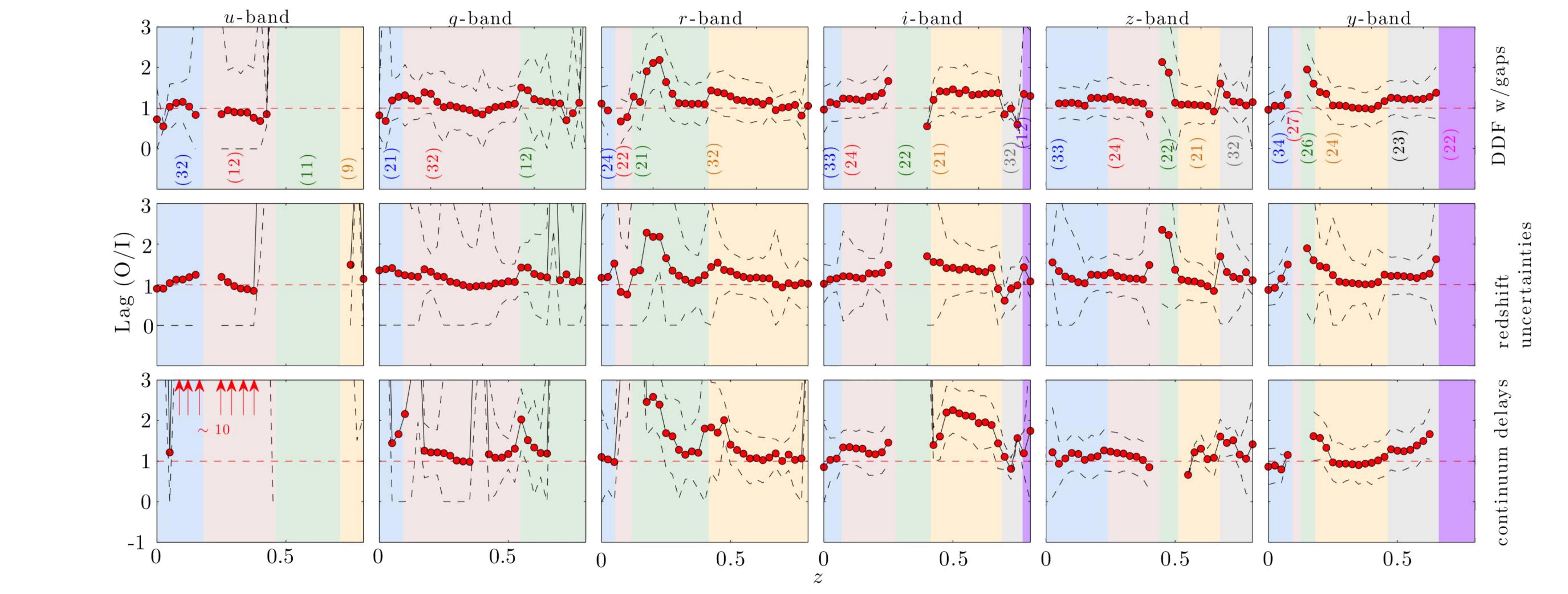}
\caption{Statistics of time-delay measurements for a $L_{\rm opt}=10^{45.5}\,{\rm erg~s^{-1}}$ source, as a function of redshift, for the different LSST bands. Red points mark the median O/I for the time delay, while the dashed line mark one standard deviation around this value (with a hard bound set at zero). The IDs of the most prominent emission features from Table 1, for a given band, and as a function of redshift, are denoted in different color bands (labeled by numbers in parentheses). All three cases shown correspond to DDF sampling with 120\,days of seasonal gaps. Upper panels' statistics are obtained from Figure \ref{zzz}. The middle row panels include the effect of finite redshift uncertainties (see text). While the median time delays are robust to finite redshift accuracies, the reliability of individual measurements at redshift ranges where filter switches take place, or when certain emission lines enter or leave the band in question, is considerably reduced. The lower panels show the effect of continuum time delays on the lag measurements.  The median time delay is sensitive to the continuum time delays, and large biases may occur depending on the choice of line-poor and line-rich bands (see e.g., the $u$-band results at $z<0.4$ where the deduced lag may exceed the input lag by as much as a factor 10; see text).}
\label{zerr}
\end{figure*}

Simulations show that more reliable solutions under sparse sampling may be obtained when the emission line contribution to the band is larger. This may occur in cases where effectively narrower photometric bands are considered or when stronger emission lines are probed (see Fig. \ref{lll} and section 4 discussing the results for the Ly$\alpha$ line). In particular, for $\alpha \to 1$ the MCF reduces to standard cross-correlation techniques with their advantages and shortcomings \citep[and references therein]{np97,wel99}.

We note that, in the limit of truly random sampling, infinite characteristic timescales of the time series exist, and the solution phase space is uniformly covered (not shown). In such cases it is practically impossible to deduce the true lag, regardless of whether additional information on $\alpha$ exists. 

\subsection{Multiple Emission Lines}

Here we include all emission lines and blends (as well as the Paschen and Balmer bumps, and dust emission) listed in Table 1, and examine the ability of the MCF algorithm to recover the line-to-continuum time delay. More specifically, we first identify the line-poor band and associate its light curve with $f_c$. We then determine the (effective) $\tau$ and $\alpha$ from the MCF for each filter combination. The solutions are then compared to the input lags and the relative contributions to the flux of the emission features that contribute the {\it most} to each of the line-rich bands. That is, whether a particular emission component contributes solely to the flux in the line-rich band, or whether its contribution is only slightly larger than that of the other emission components contributing to the same band, does not alter our interpretation of the signal. 

Results for individual objects having a luminosity of  $10^{45.5}\,{\rm erg~s^{-1}}$, over the redshift range $0<z<3$, assuming DDF sampling with seasonal gaps, are shown in Figure \ref{zzz} for the different bands. For the chosen luminosity, the algorithm is able to determine a lag up to $z\lesssim 2$. Specifically, as the S/N decreases, the recovered time-delay distribution extends to shorter recovered lags whence becoming less reliable. Non-physical solutions may be discarded, as before, using prior information on $\alpha$, leading to solutions which hover around the input value, with a typical scatter of 50\% (see also upper panels of figure \ref{zerr}). 

Interestingly, some filter combinations give rise to more biased results. For example, time delay measurements in the $r$ band for $z\sim 0.2$ objects (here, the $u$ band is the line-poor band), show a factor $\sim 2$ overestimation of the lag (see Figs. \ref{zzz},\ref{zerr}). The primary reason for that has to do with the fact that, at such redshifts, \ion{Fe}{2} and H$\beta$ have comparable contributions to the flux in the $r$-band, with the former transition being characterized by longer time-delays \citep[see our Table 1]{bar13,cr13}. We also note that, at $z\sim 0.2$, the contribution of the emission lines to the $r$ band is only slightly larger than that to the $u$ band (see Fig. \ref{frac}) and caution is advised when interpreting the signal. 

It is also interesting to note the increased scatter in the recovered lags at redshifts where the identity of the line-poor filter is replaced, or around redshift intervals where the line-rich band gradually becomes line poor, hence the BLR signal weakens (e.g., the case of the $u$ band at $z\lesssim 0.2$; Fig. \ref{zerr}).

\subsubsection{Redshift uncertainties}

Redshift inaccuracies may lead to an erroneous identification of the line-poor and line-rich bands, and to poor estimation of the relative contribution of emission lines to the flux, hence to an improper interpretation of the signal. We note that typical redshift uncertainties in photometric surveys are of the order of a few percent, and incorporated such uncertainties in our simulations pipeline (a Gaussian distribution with a standard deviation, $\delta z =0.05$  was assumed, and a hard lower bound of $z=0$ enforced).

Figure \ref{zerr} (middle panels) portrays the effect of redshift uncertainties, where the standard deviation of the recovered lag distribution rapidly increases at band-specific redshift ranges. As expected, such intervals correspond to redshifts where the identity of the most prominent emission line at any given band is changed over narrow redshift intervals, or in cases where the identity of the line-poor band is switched. Most importantly, while individual measurements are less reliable at such redshift intervals, the median of the recovered lag distribution does not show significant deviation from the ideal case where the redshift is precisely known. Clearly, provided systematic effects are well controlled, large number statistics wins.

Lastly, we mention a subtle effect having to do with the emission line contribution to the band across filter edges. Specifically, for relatively boxy filter throughput curves it is possible for only one of the extended line wings to contribute to the flux in the band. As has been shown in several previous works, the time delay associated with emission line wings could differ from that which characterizes the bulk of the emission line \citep[and references therein]{gr13}. Such effects could lead to small biases in the median lag close to certain redshift intervals. However, under most circumstances pertaining to broadband data, the emission line contribution at the filter edge will be overwhelmed by that of another transition close to the peak sensitivity curve of the band. The full treatment of such cases is beyond the scope of the present work.

\subsection{Continuum Time Delays}

The presence of time delays between the continuum emission in different wavebands (equation \ref{tjk}) could considerably bias line-to-continuum time-delay measurements, as already discussed in \citet{cd11}. Figure \ref{zerr} (bottom panels) shows this effect when considering a quasar with an optical luminosity of $10^{45.5}\,{\rm erg~s^{-1}}$ over a broad redshift range.  

We find that, for much of the parameter space covered by our simulations, the effect of continuum time delays on the deduced (line-to-continuum) lag is modest. Nevertheless, there are particular filter combinations and redshift intervals where the deduced (median) lag can considerably deviate from the input value, which cannot be corrected by using priors on $\alpha$. For example, the emission-line time-delays associated with the $u$-band appear to be highly biased (by an order of magnitude) compared to  the input delay for $z<0.4$ sources. This is due to the fact that the $i$ band is used as the line-poor band at this redshift interval, and due to the continua time delay between the $i$ band and the $u$ band being $\sim10$\,days in our model, i.e., $\sim 6\%$ of the line-to-continuum time delay in this case (see also \citealt{cd11}). A similar bias, although somewhat smaller in amplitude, is  noticeable when considering the $g$ and $r$ bands at $z\sim 0.2$. 

There are two straightforward ways to mitigate the aforementioned complications: (1) to use a line-poor band that is as adjacent as possible to the line-rich band probed, so that inter-band continuum time delays are minimized (equation \ref{tjk}), or (2) to use a more comprehensive formalism, which is able to handle situations where two lagging emission components are present in the problem. The full treatment of such situations is naturally more complicated and may also lead to degenerate results \citep{cz13}. 

Lastly, we mention the possibility of non-negligible diffuse continuum emission from the BLR \citep[see their Figs. 2, 4]{kor01}. While the quantitative treatment of this scenario is beyond the scope of the present work, and is subject to model uncertainties, at the basic level, the problem is akin to that of the continua time-delays. Specifically, one should seek line-poor and line-rich filter combinations for which diffuse continuum emission from the BLR has similar characteristics (e.g., lag and relative contribution to the flux).

\section{Discussion}

\begin{figure}
\epsscale{1.2}
\plotone{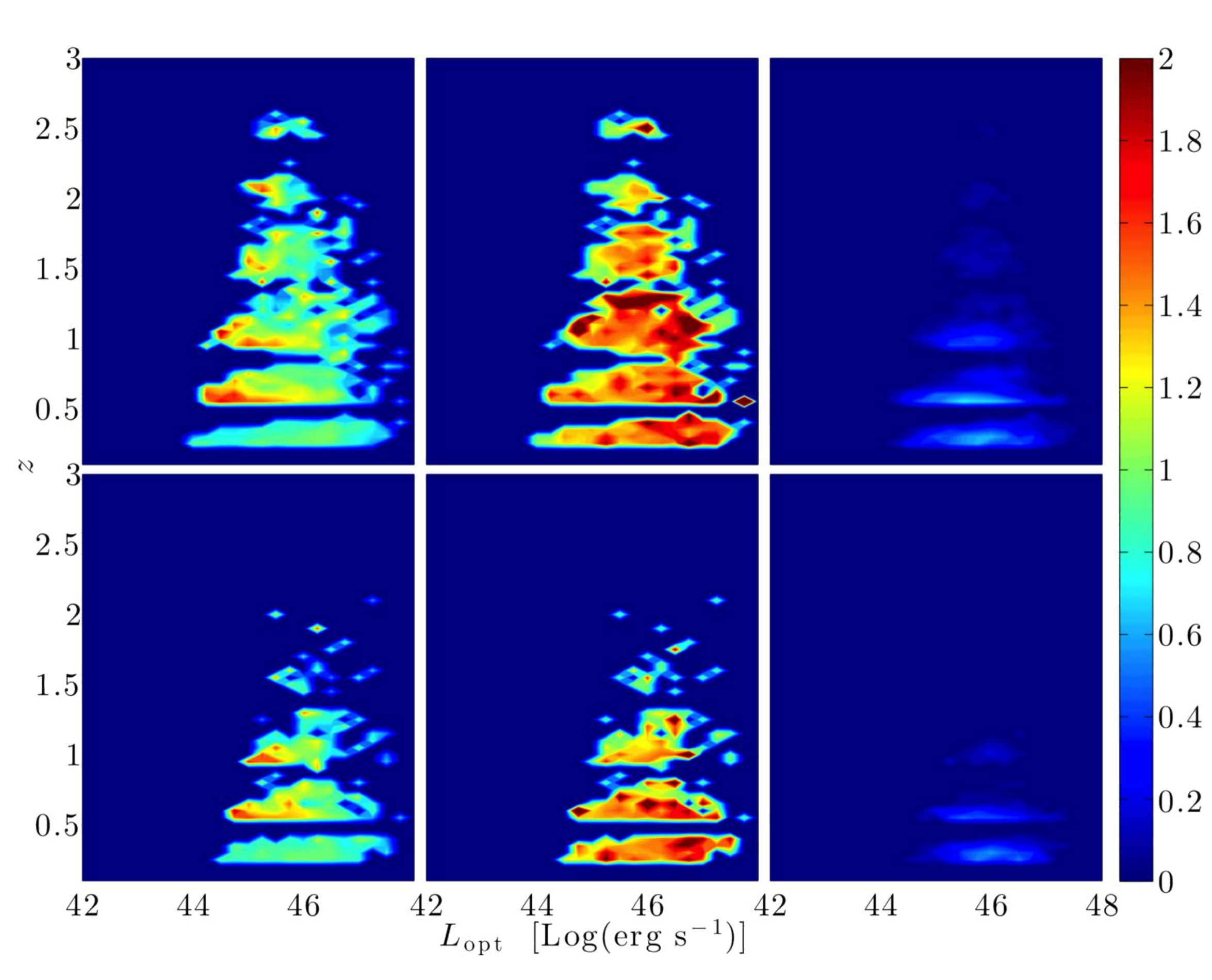}
\caption{Lag determination for the \ion{Mg}{2}\,$\lambda 2799$ emission line in the luminosity-redshift plane. The top row corresponds to a 60\,sec exposure (obtained by combining 4 exposures of 15\,sec each, taken over an interval of 30\,minutes), while the bottom row assumes 15\,sec-exposure visits. Only objects for which $\alpha$ is recovered to within $\pm40\%$ of the input value are considered in the presented statistics, which results in only a fraction of all objects being used. The blue-to-red colors (see color bar) mark the median output-to-input (O/I) lag ratio (left column) in each redshift-luminosity bin, the standard deviation (shown in the middle column as $10^{\rm STD[Log(O/I)]}$), and the fraction of objects that pass our $\alpha$-selection criterion (see text; right column). DDF sampling is assumed in all cases.}
\label{Lz_mg2}
\end{figure}

\begin{figure*}
\epsscale{1.17}
\plottwo{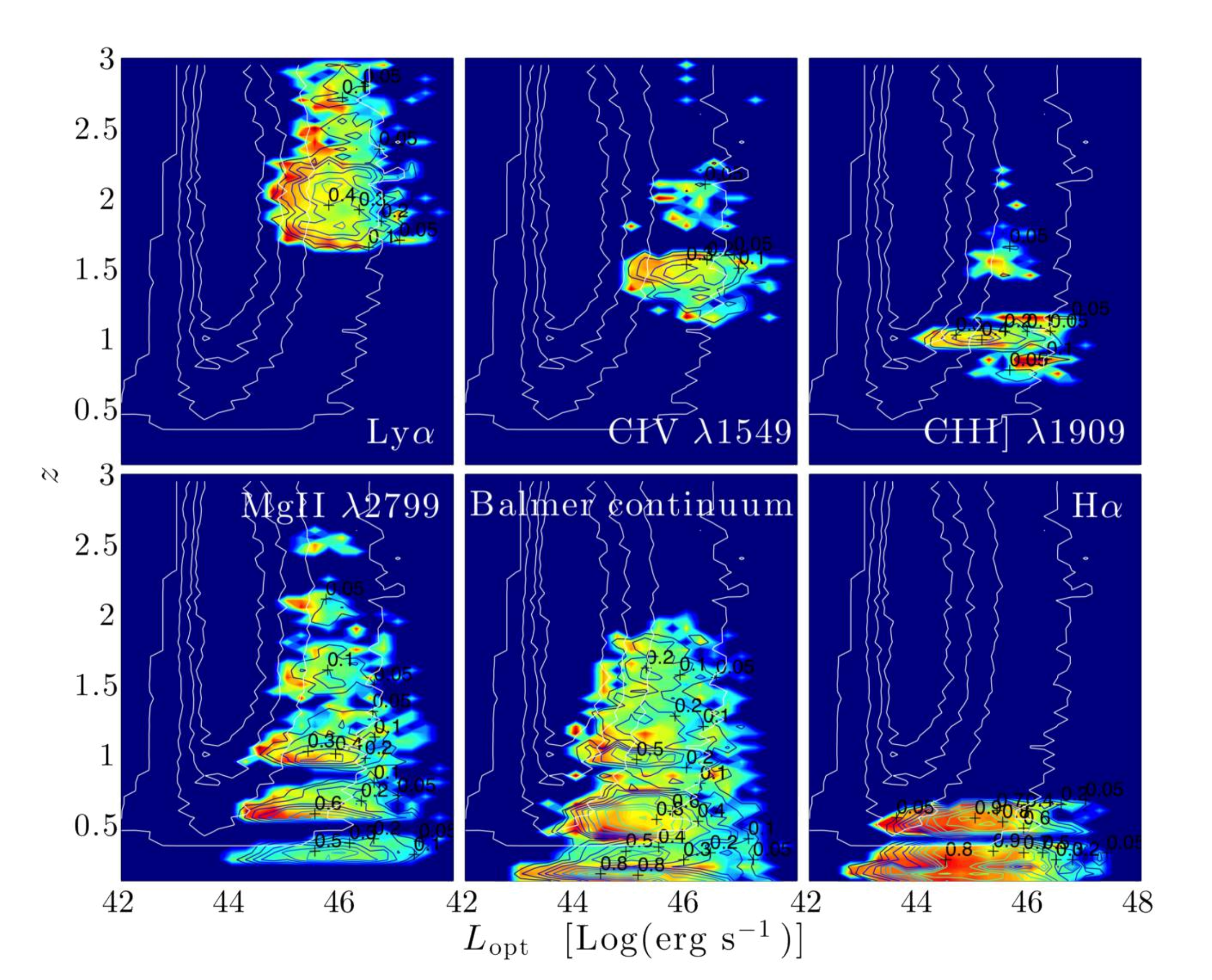}{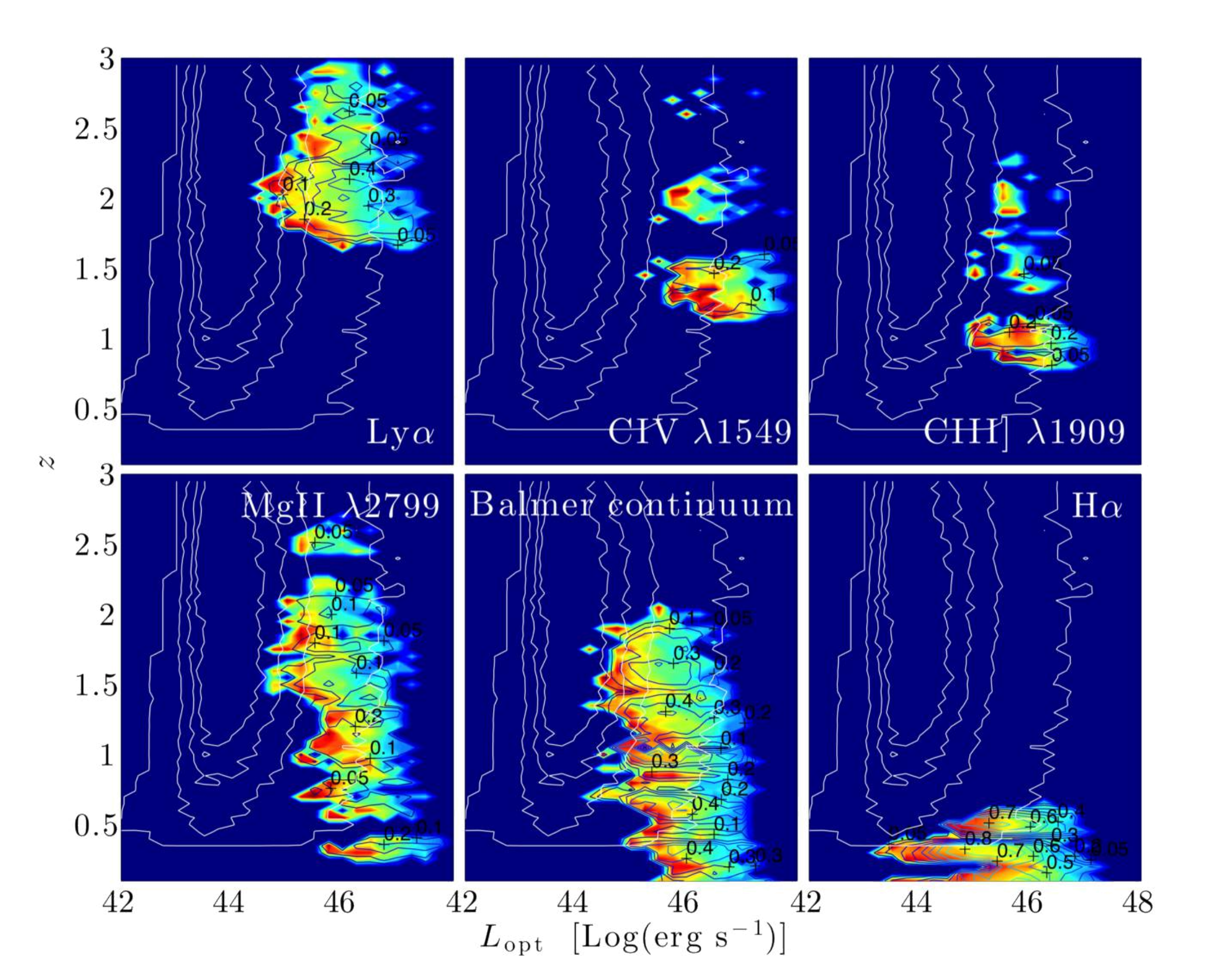}
\caption{Lag determination for prominent BLR emission features in the luminosity-redshift plane. Colored regions mark the value of the median ratio of the output-to-input (O/I) time delay (the colorbar of figure \ref{Lz_mg2} applies here: green colors mark values of unity while blue/dark-red shades indicate $-/+0.3$\,dex deviations). Only objects for which $\alpha$ is recovered to within $\pm40\%$ of the input value are considered in the presented statistics, which results in only a fraction of all objects being used (this is marked as dark blue contours with values denoting fraction levels). Also shown in white contours are regions in the luminosity-redshift plane including 60\%, 70\%, 80\% ,90\% and 99.99\% of all LSST-selected quasars (this does not include faint AGN). {\it Left panels:} Results for DDF sampling with 4 months' seasonal gaps show that time-lag measurements are S/N bounded at low luminosities and limited by the total lifetime of the LSST survey at high-luminosity and redshifts. {\it Right panels:} Results for UNS, which characterize the bulk of LSST sources, are bounded at the low-luminosity end by relatively poor sampling. Generally, reverberation mapping is feasible in roughly the brightest $10\%$ of the quasar population, with DDF sampling allowing to recover shorter BLR timescales in less luminous sources, especially at low-$z$.}
\label{Lz}
\end{figure*}

We have shown that photometric surveys with quasi-regular sampling in several bands, and with characteristics similar to the LSST, can be used to measure the line-to-continuum time-delay in major emission lines, line blends, and non-ionizing continua from the BLR.  Lag measurements show enhanced scatter when the time delay is a fair fraction of the survey lifetime, or when the lag is comparable to the sampling period. Nevertheless, even in such cases, it is possible to reduce the scatter and reach statistically robust results if additional constraints on the value of $\alpha$ (e.g., from independent spectroscopic measurements, or from knowledge of the quasar population as a whole) are incorporated in the MCF analysis. Probing line-to-continuum time-delays on sub-sampling timescales is unreliable, as photometry cannot disentangle line and continuum light curves on short timescales; an exception to this rule is in cases where the lagging signal dominates the flux in the band \citep{cz13,c13}, and provided the light curves are well sampled. 

Time series whose sampling pattern exhibits many different characteristic timescales lead to considerable scatter in individual lag measurements, especially on timescales comparable to the sampling periods and their harmonics. Here too, it is possible to reject non-physical solutions by employing priors on $\alpha$ so that a statistical measurement of the median lag, in properly defined quasar samples, provides a good estimator for the true lag. The degree to which such filtering can be effectively used depends on the strength of the emission feature probed. Generally, as with spectroscopic reverberation mapping campaigns, regular sampling should be sought as one attempts to reduce the number of sampling timescales in the problem.

Good redshift determination leads to more robust lag measurement on a case-by-case basis since preferable filter combinations may be reliably selected, and meaningful priors on $\alpha$ may be set.  This also exemplifies the advantage of having follow-up (single-epoch) spectroscopy which can secure the object identification and precisely determine its redshift. Moreover, spectroscopic follow-up can determine the velocity dispersion of the relevant emission lines, thereby providing critical information required for SMBH mass estimates.

We find that, even under ideal observing conditions, the recovered median lag may overestimate the true lag, i.e. the centroid of the line transfer function,  by $\lesssim 20$\%. This is a direct consequence of the inability of photometric data to disentangle line and continuum light curves on short timescales, whose sum makes up the total signal. The exact value of the bias depends primarily on the line transfer function, which is rather loosely constrained by observations: for transfer functions with a large amplitude at zero time delays (e.g., when line emissivity close to our sightline is considerable, as in edge-on configurations) the bias is more significant. Consequently, the bias is smaller for face-on systems, which may be more relevant to type-I quasars \citep{mai01}. This implies that some statistical information concerning the line transfer function may be obtained using photometric means. 

Our simulations indicate that photometric reverberation mapping is especially advantageous when large samples are concerned, as the median time delay is less susceptible to sampling-induced noise, redshift uncertainties, and the underlying properties of the power-density spectrum of the quasar.  Therefore, good control of systematic effects is essential, especially as far as redshift determination is concerned. Provided large enough samples exist, statistical averaging would lead to BLR size-luminosity relations with unprecedented accuracy for various emission lines, and over a broad luminosity and redshift ranges (see also \citealt{cd11,zu13}, and below). 

It is worthwhile to consider more specific predictions for the LSST. We first consider the case where the signal per visit is obtained by combining the data of four consecutive (to within $\sim$30\,minutes)  exposures of 15\,sec each, which should characterize the majority of LSST data\footnote{See \url{http://www.lsst.org/files/docs/sciencebook/SB_2.pdf}}. Thus, we effectively neglect quasar variations on $\sim$hour timescales, which is reasonable given the red power-density spectra of quasars, the suppressed variability of luminous high-$z$ sources (Fig. \ref{var}), and the considerably longer timescales associated with the BLR.  

Figure \ref{Lz_mg2} divides the luminosity-redshift space into small [$\delta {\log}(L_{\rm opt})=0.25,\delta z=0.05$] segments, each typically having $\lesssim 10^3$ simulated sources, and shows the statistical properties of the obtained lag solutions for \ion{Mg}{2}\,$\lambda 2799$. The time-delay statistics includes objects for which the recovered $\alpha$ is within $\pm 40$\% of the input value (section 3).  The relevant parameter space for measuring the size of the \ion{Mg}{2}\,$\lambda 2799$ emission region is well-defined and forms an envelope that is bounded at low-luminosities by minimal S/N requirements, and by the finite duration of the survey at the high-luminosity end. Within this envelope the recovered lag is within $\pm0.3$\,dex of the input value, with a median value of $\simeq 0.9$ over the relevant phase space. In particular, several trends are observed: in regimes of effective low-S/N (either low source flux and/or small contribution of the emission line to the flux in the band), there is a tendency for a biased lag measurement, by up to 60\%; c.f., the second column of Figure \ref{zerr}. There is also a tendency to somewhat underestimate the lag, by typically 30\%, in cases where it is comparable to the lifetime of the experiment. Discarding those regions near the envelope's rims, typical median lag determination is at the $\sim \pm 20$\% level for the \ion{Mg}{2} line out to $z\sim 2.5$. Taking into account the quasar luminosity function and the LSST selection criteria, some $5\times 10^4$ lags may be determined over the LSST lifetime (see also Table 2) suggesting that \ion{Mg}{2} may have a crucial role in constructing more reliable single-epoch BH mass estimations by cross-calibrating scaling relations used at different redshift ranges.

Our calculations imply that the expected scatter in individual \ion{Mg}{2}\,$\lambda 2799$ lag measurements is, typically, $\sim 40$\%, with a tendency for an increased scatter under low S/N conditions (see the middle panels of Fig. \ref{Lz_mg2}).  As noted above, these statistics were obtained after screening against erroneous $\alpha$-solutions that, typically, results in only $\sim$ 10\% of the objects being used (right panel of Fig. \ref{Lz_mg2}). Nevertheless, there are regions in the parameter space ($z<1.2$ and $45<{\rm log}(L_{\rm opt})<46$) that are characterized by good S/N, and well-sampled light curves with respect to the BLR extent, for which $\gtrsim 30$\% of the sources lead to robust solutions. We note that it may be possible to further narrow the scatter in individual measurements by testing the robustness of the solutions \citep{cz13}. This, however, is beyond the scope of the present work due to the prohibitively long computation time involved.

Reducing the exposure time per visit to 15\,sec instead of 60\,sec, shrinks the parameter space over which reliable \ion{Mg}{2} lag measurements are obtained while maintaining similar lag statistics in regions of the parameter space where such measurements are possible. Specifically, with the S/N reduced by a factor 2, \ion{Mg}{2}\ lag measurements are feasible up to redshift of $\sim 1.7$, instead of $2.5$, and only for the brighter sources, which reduces the number of reliable lag measurements by an order of magnitude for this transition (Table 2).

\begin{figure}
\epsscale{1.2}
\plotone{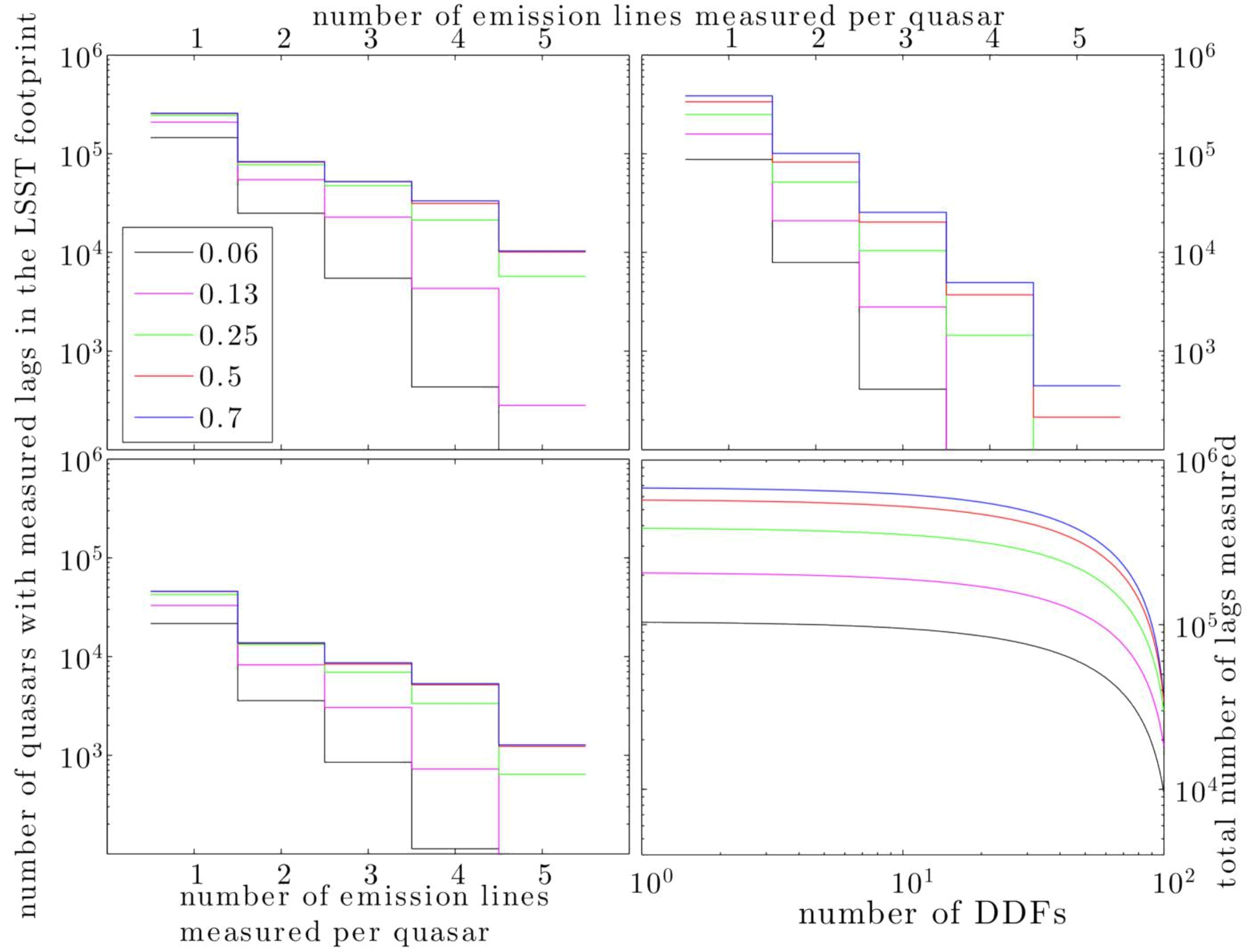}
\caption{Lag determination statistics with the LSST. Upper left panel shows the number of objects within the LSST footprint for which a given number of emission-line time delays may be simultaneously determined (using six filters a maximum of five lags are measurable using the formalism adopted here).  A total exposure time of 60\,sec per visit is assumed and DDF sampling is considered. Different colors correspond to different lag measurement accuracies; for example, the blue line includes all objects for which the ratio between the absolute difference of the measured to input lag, and the input lag is $<10^{0.7}$ (see legend). Upper-right panel is for UNS sampling, while the lower-left panel assumes DDF sampling but with a 15\,sec exposure time per visit. The bottom-right panel is the total number of time-delay measurements, as a function of the number of DDFs surveyed (see text).}
\label{ddfuns}
\end{figure}

Figure \ref{Lz} shows an analysis similar to the above for several additional major emission lines in quasar spectra, and examines some of their statistical properties using DDF and UNS samplings (60\,sec exposure times are assumed throughout). Clearly, there exists a non-negligible region in the redshift-luminosity plane in which time delays may be determined with good accuracy. For example, in low-luminosity, low-$z$ quasars, the lag-luminosity relation for H$\alpha$ may be probed over $\sim 2$ orders of magnitude in luminosity with a total expected number of H$\alpha$ lag measurements of $\gtrsim 10^4$ (Table 2). For the \ion{C}{4}\,$\lambda 1549$ emission line, the parameter range probed extends to lower luminosities than probed by \citet{kas07}, and is dominated by radio-quiet objects rather than radio loud ones, as in their work. The small blue bump (Balmer continuum) has a similar role to the \ion{Mg}{2} line in the sense that its lag may be quantified over a broad range of quasar luminosity and redshift. Obviously, this spectral feature is of limited use for directly estimating BH masses as there is no kinematic information associated with it. Interestingly, we find that Ly$\alpha$ lag measurements at the peak of quasar activity are facilitated by the effectively narrow and spiky throughput curve of the $u$ band, and are feasible out to $z\sim3$.

\begin{table}
\begin{center}
\caption{Lag measurement statistics in the LSST footprint}
\begin{tabular}{llll}
\tableline
 &  UNS & DDF$^\dagger$  &  DDF$^\dagger$ \\
ID  &   (60\,sec) &(60\,sec) & (15\,sec) \\          
\tableline
Ly $\alpha$ 	& 64000 	& 95000 	& 7000 	\\
\ion{Si}{4}\,$\lambda 1397$		&	150	& 300 	& 30		\\
\ion{C}{4}\,$\lambda 1549$		&	8000	&16000	& 1200	\\
\ion{C}{3}$]$\,$\lambda 1909$	&	21000	&45000 	& 6500	\\
\ion{Mg}{2}\,$\lambda 2799$	&	46000	&48000	&4300 	\\
H$\gamma$	&	2000	& 6000	&300	 \\
\ion{Fe}{2}\,$\lambda 4564^\ddag$		&9400	&16000	&900	 \\
H$\beta$		&	47000	&110000	&19000	\\
\ion{Fe}{2}\,$\lambda 5305^\ddag$		&	6500	&27000	&4800	\\
\ion{He}{1}\,$\lambda 5877$		&		0	&700	&40 \\
H$\alpha$		&	28000	&73000	&25000 \\
Balmer cont. 	&	310000	&310000	&50000 \\
Paschen cont.	&	14000	&15000	&4000 \\
\tableline
Total for all BLR features:  &560000	&760000	&120000 \\

\tableline
\end{tabular}
\end{center}
Number of time-delays measured for individual emission features in quasars within the LSST footprint for DDF (using 15\,sec and 60\,sec exposure times) and UNS sampling (60\,sec exposure time). Features for which $>100$ lag measurements may be obtained are included in the table, and measurements for which the measured lag agrees to better than 0.5\,dex with the input value. Typical uncertainties on the quoted figures are at the 10\% level given the model assumptions (but may be of order unity for those transitions with $\lesssim 100$ detection statistics).\\
$^\dagger$ Full coverage of the LSST footprint by DDFs exceeds the LSST resources (see text). \\
$^\ddag$ Approximate wavelengths are quoted for blends (Table 1).
\end{table}

The fact that the lags of different emission lines may be determined in overlapping luminosity-redshift ranges means that a more reliable BLR-size ladder (in analogy with the cosmological distance ladder) may be obtained using LSST, allowing the cross calibration of different prescriptions for SMBH mass estimation over cosmic time. In this respect, it is interesting to note that for an exponentially-small (but finite) fraction of all quasars, simultaneous lag determination for several transitions may be possible. The number statistics in this case is less secure and may depend on the exponential tails of currently poorly-determined quasar property distributions (e.g., the fraction of objects with extreme variability amplitude). Current predictions for the statistics of multi time-lag measurements within the LSST footprint for different levels of accuracy, and for UNS and DDF sampling (for the latter either 15\,sec or 60\,sec exposure times are considered) are shown in Figure \ref{ddfuns}.  Clearly, an exponentially small (but finite) number of sources will have up to 5 time-delays simultaneously determined. The number of such sources is very sensitive to sampling: e.g., the number of quasars with five lag measurements drops by more than an order of magnitude when switching from DDF to UNS sampling. S/N has a smaller effect on the relative number of objects with multi time-delay measurements. 

Lastly, we consider the total number of reliable time-lag detections as a function of the number of DDFs covered assuming each field consumes 1\% of the LSST resources\footnote{Assuming our definition of a DDF, each field, observed daily, will require four 15\,sec exposures, each with an overhead of 2\,sec of reading time, in six bands. Slewing between adjacent fields will take additional 5\,sec. Together, this amounts to $\sim 400$\,sec per DDF, which is of order 1\% of a full observing night.}. The total sky coverage in $N$ DDFs is then $\simeq 9N$\,sq-deg, and the remaining UNS coverage is then $\simeq 2\times 10^4(1-N/100)$\,sq-deg. The total number of time lag measurements as a function of $N$ is shown in the bottom-right panel of Figure \ref{ddfuns} demonstrating that sheer number statistics prefers UNS over DDF sampling with up to $\sim 500,000$ time delay measurements possible for full UNS coverage of the sky.  Note, however, that the sparser UNS will result in the loss of short time-delay information concerning, e.g., accretion disks or high-ionization optical emission lines, such as \ion{He}{1}\,$\lambda$5877 (see Table 2), especially in low-$z$, low-luminosity sources. 

\section{Summary}

We show that large time-domain photometric surveys, such as the LSST, are expected to transform the field of AGN reverberation mapping by increasing the number of objects with measured time-delays by several orders of magnitude and by including sources at the epoch of peak quasar activity, at $z\sim 2$. Specifically, the LSST is expected to yield a total of $\gtrsim 10^5$ time-delay measurements including all major emission lines, blends, and pseudo-continuum features out to $z\sim 3$. For a given exposure time per visit and finite survey resources, covering a larger area of the sky at the expense of sampling would result in additional time-lag measurements while inhibiting the study of short time-scale phenomena. Spectroscopic follow-up of photometric quasars will lead to improved BLR lag-determination via the incorporation of priors in the MCF analysis. It  will also facilitate BH mass estimations thereby improving our understanding of black hole demography at high-$z$ with implications for galaxy evolution, black hole seeds, and gravitational wave detection.

\acknowledgements 

We thank E. Behar, W. N. Brandt, L. R. Jones, M. Juri\'c, S. Kaspi, and G. T. Richards for fruitful discussions and good advice, and an anonymous referee for valuable comments. This research has been supported in part by a FP7/IRG PIRG-GA-2009-256434 and by grant 927/11 from the Israeli Science Foundation, and the Jack Adler Foundation. Research by A.J.B. is supported by NSF grant AST-1108835. S. E. R. is supported at the Technion by the Zeff Fellowship. Simulations carried out in this study were performed on the TAMNUN high-performance computer cluster funded by the Minerva Foundation and the Russel Berrie Nanotechnology Institute at the Technion; we thank E. Behar for facilitating access to this computer.

\end{document}